\documentclass[a4paper]{article}
\usepackage{graphics}
\usepackage{shonan}
\usepackage{graphicx}
\usepackage{hyperref}
\usepackage[normalem]{ulem}
\usepackage{pdflscape}
\usepackage{amsmath}
\usepackage{amsfonts}

\usepackage{tikz}
\usetikzlibrary{fit, calc, decorations.markings}

\begin{document}

\SHONANno{189}
\SHONANtitle{Advancing Visual Computing\\in Materials Science}
\SHONANauthor{%
Christoph Heinzl\\
Renata Georgia Raidou\\
Kristi Potter\\
Yuriko Takeshima\\
Mike Kirby\\
Guillermo Requena}
\SHONANdate{May 13--17, 2024}
\SHONANmakecover

\title{\textbf{Advancing Visual Computing\\in Materials Science}\footnote{\url{https://shonan.nii.ac.jp/seminars/189/}}}
\author{
Christoph Heinzl\thanks{University of Passau, Germany, \href{mailto:christoph.heinzl@uni-passau.de}{christoph.heinzl@uni-passau.de}}, 
Renata Georgia Raidou\thanks{TU Wien, Austria, \href{mailto:rraidou@cg.tuwien.ac.at}{rraidou@cg.tuwien.ac.at}},\\
Kristi Potter\thanks{National Renewable Energy Laboratory, USA, \href{mailto:Kristi.Potter@nrel.gov}{Kristi.Potter@nrel.gov}},
Yuriko Takeshima\thanks{Tokyo University of Technology, Japan, \href{mailto:takeshimayrk@stf.teu.ac.jp}{takeshimayrk@stf.teu.ac.jp}},\\
Mike Kirby\thanks{University of Utah, USA,\href{mailto:kirby@cs.utah.edu}{kirby@cs.utah.edu}},
Guillermo Requena\thanks{RWTH Aachen University / DLR, Germany, \href{mailto:Guillermo.Requena@dlr.de}{Guillermo.Requena@dlr.de}}
}
\date{May 13--17, 2024}
\maketitle

\begin{figure}[h!]
    \centering
    \includegraphics[width=.85\linewidth]{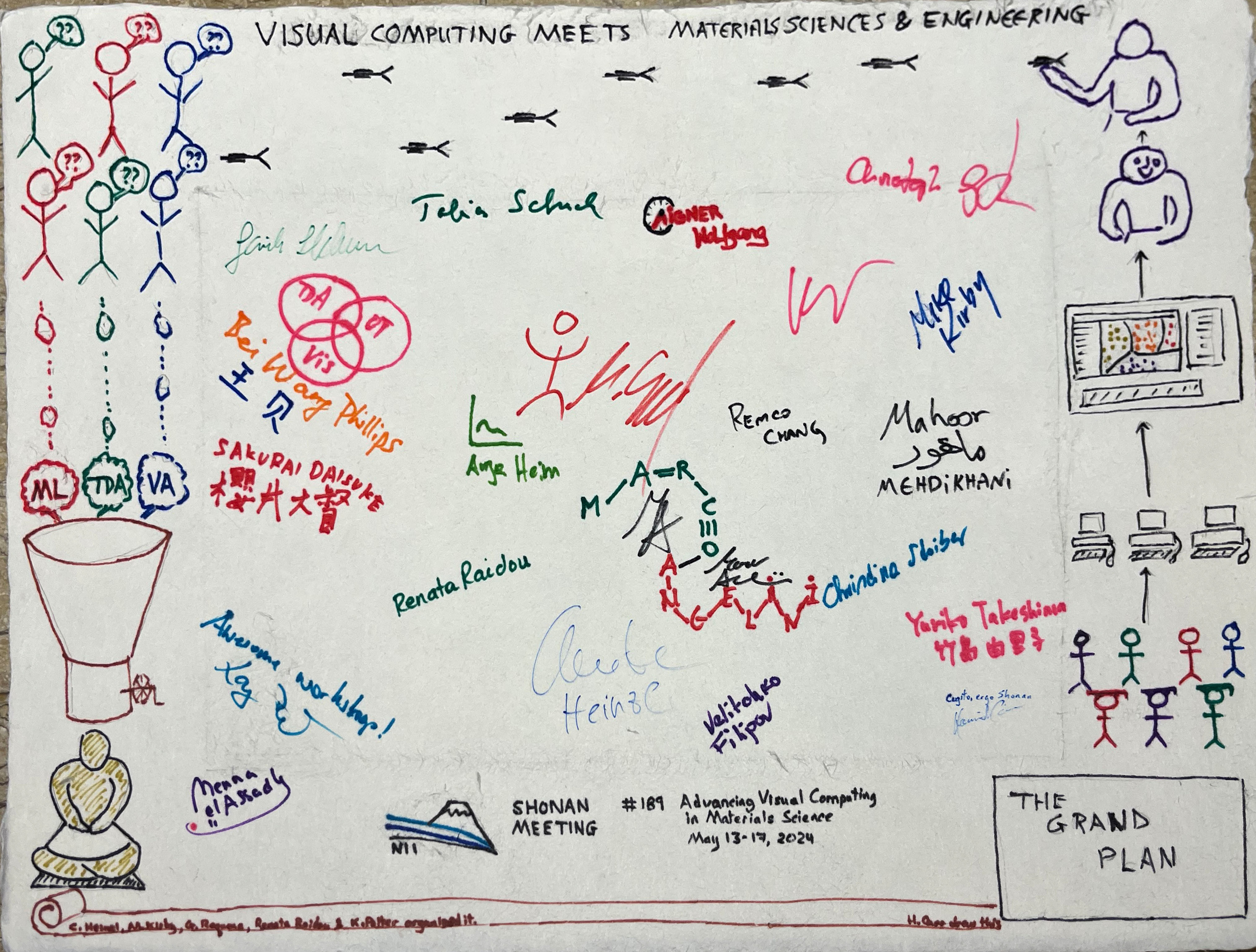}
    \label{fig:teaser}
\end{figure}

\begin{abstract}
Materials science has a significant impact on society and its quality of life---for example, through the development of safer, healthier, more durable, more economical, environmentally friendly, and sustainable materials. Visual computing in materials science integrates computer science disciplines from image processing, visualization, computer graphics, pattern recognition, computer vision, virtual and augmented reality, machine learning, to human-computer interaction, in order to support the acquisition, analysis, and synthesis of (visual) materials science data with computer resources. Therefore, visual computing may provide fundamentally new insights into materials science problems by facilitating the understanding, discovery, design, and usage of complex material systems~\cite{heinzl_star_2017}.

We consider this seminar as a follow-up of the Dagstuhl Seminar 19151 Visual Computing in Materials Sciences~\cite{heinzl_et_al:DagRep.9.4.1}, held in April 2019. Since then, the field has kept evolving and many novel challenges have emerged, with regard to more \textit{traditional topics} in visual computing, such as topology analysis or image processing and analysis, to \textit{recently emerging topics}, such as uncertainty and ensemble analysis, and to the integration of \textit{new research disciplines and exploratory technologies}, such machine learning and immersive analytics. The 2019 seminar aimed at setting the fundamentals of the field and stimulating cooperation with domain experts. With the current seminar, we target to strengthen and extend the collaboration between the domains of visual computing and materials science (and across visual computing disciplines), by foreseeing challenges and identifying novel directions of interdisciplinary work. 

During this \emph{5-day seminar}, which was delayed by more than 2,5 years due to COVID19, we brought visual computing and visualization experts from academia, research centers, and industry together with domain experts, to uncover the overlaps of visual computing and materials science and to discover yet-unsolved challenges, on which we can collaborate to achieve a higher societal impact. We see this as a great opportunity to cover a range of relevant and contemporary topics, and to identify the most significant directions for future work. By organizing the seminar at the Shonan Village Center, we bring together top-notch researchers from the entire world and especially aim at bridging new collaborations with Asian institutions. 

\end{abstract}

\normalsize
\section*{Background and Introduction}
In our seminar, our goal is to bring together researchers working in close proximity to the interdisciplinary domain of visual computing in materials science, in order to identify current, unsolved challenges and to discuss opportunities for future collaborations.
This cannot be done easily during conferences, as there are no dedicated venues, where both visual computing and material science experts participate, and can discuss in an organized and structured manner.
Topic-wise, our seminar aims at covering the most relevant trends within visual computing and linking them to current and future challenges in materials science. 
We, therefore, aim to discuss topics that address analytical processes through their entire spectrum---from \textit{exploration to decision making}.
These topics span from \textit{traditional directions}, such as topology analysis, to recently \textit{emerging directions}, such as uncertainty and ensemble analysis, and to \textit{new technologies}, such as machine learning and immersive analytics. 

We expect that our seminar will contribute to new interesting paths of research and will significantly advance the domain of visual computing in materials science in many ways, which is simply impossible without this seminar.
Despite the specific objectives of the application domain, we anticipate that our joint effort might be useful and applicable to other domains of interdisciplinary work.

We have identified three main directions, which have not been tackled in depth or at all at the previous Dagstuhl seminar and we would like to address them at our Shonan seminar. These are: \textit{Topology visualization}, \textit{Integrated visual analysis}, and \textit{Interpretability and decision making}. We have categorized these directions (and their respective topics) based on two dimensions: \emph{traditional research vs. new trends} and \emph{exploration vs. decision making}. 
We summarize our categorization in Figure~\ref{fig:overviewfigure} below.
\vspace{15pt}

\begin{figure}[h!]
    \includegraphics[width=\textwidth]{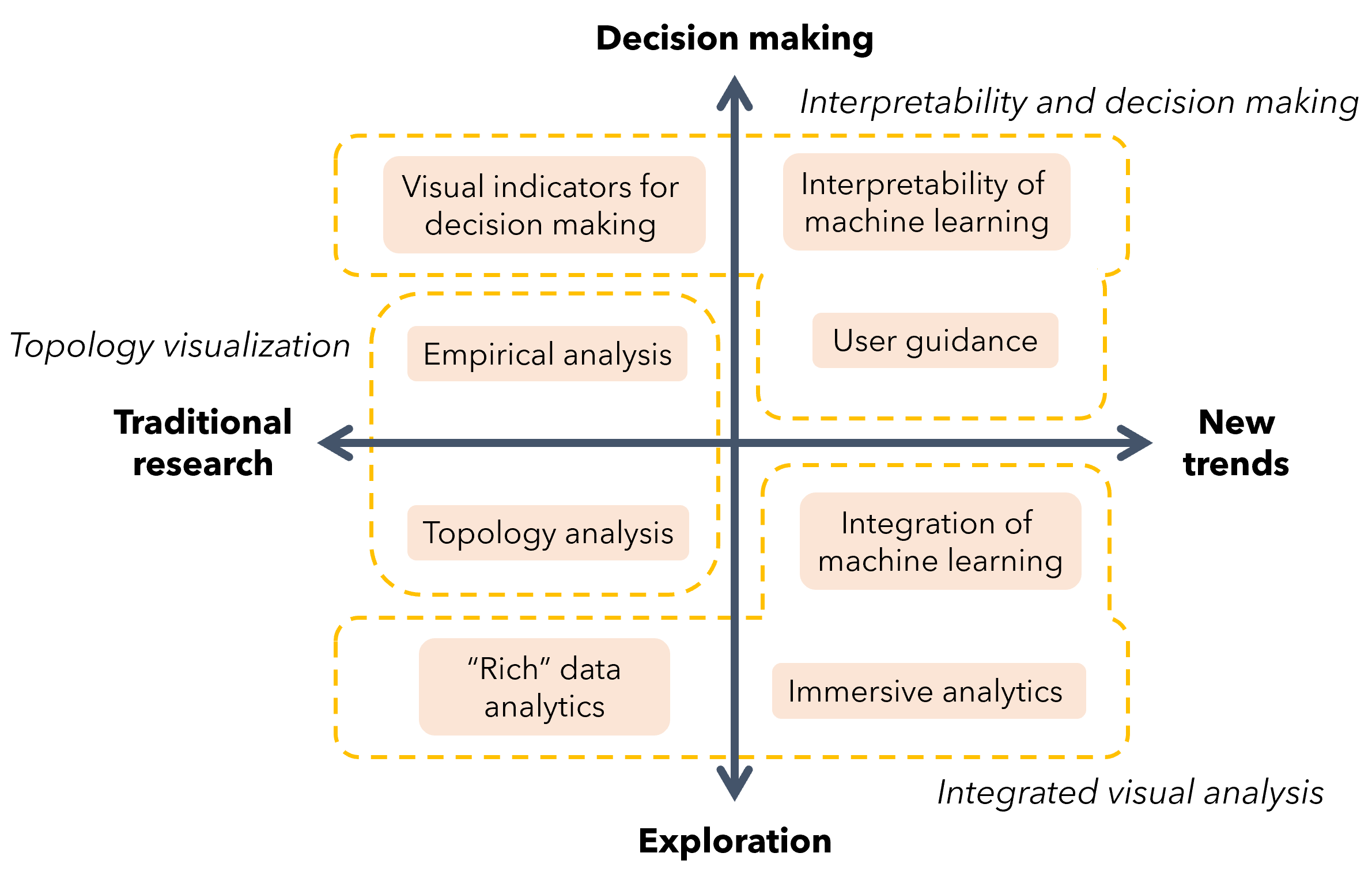}
    \caption{The topics covered in our meeting.}
    \label{fig:overviewfigure}
\end{figure}

\noindent \textbf{Topology-based visualization for materials science:} To effectively visualize large-scale and complex datasets in materials science, a data-centric approach based on the structural feature of the target dataset is indispensable. In particular, topology is an effective structure for analysis in scientific applications. In this seminar, we would like to discuss two main topics from the theoretical and practical perspectives of topology-based visualization:
\begin{itemize}
    \item \textbf{Topological analysis of materials:} We target a discussion on the significance of the topology of inherent features, such as complex internal structures (e.g., pore or crack network analysis, permeability analysis) in materials science. We mainly consider the physical meaning of the topology in materials science and also discuss the seeds and needs for topology-based visualization.
    \item \textbf{Empirical analysis of persistency:} The topological analysis of practical datasets extracts many critical points, as the target dataset involves small amplitude noise in some cases. Such minor critical points might be hiding an important global structure within the dataset, and we should remove them appropriately. We will discuss strategies to evaluate their importance and methods to reduce them effectively.
 \end{itemize}
 
\noindent \textbf{Integrated visual analysis} is of core interest in materials science. Aside from the data analysis and data exploration aspect, quantitative data visualization is also of high relevance and highly challenging in the materials science domain, e.g., for precise modeling and simulation of material systems. Features of interest need to be extracted in the data over spatial, temporal, or even higher dimensional data domains and are crucial to refine material simulations, as well as predicting material properties. In this seminar, we target to discuss three novel streams coming in, which show high potential in this domain:
\begin{itemize}
    \item \textbf{``Rich'' data analytics:} Visual analytics has become particularly popular within the domain of materials science. We target a discussion on the best practices and challenges of designing, developing, and employing scalable solutions for the quantification, exploration, and analysis of ``rich'' materials science data, i.e., multi-dimensional, multi-faceted, complex, and possibly time-varying and/or multi-modal data. Within this discussion, we will incorporate topics from the domains of ensemble and comparative visualization. 
    \item \textbf{Integration of machine learning:} Deep Convolutional Neural Networks (CNNs) and corresponding classifiers have been able to identify and classify features with very high probabilities in various application areas. Current approaches have successfully trained and applied Deep CNNs to segmentation, for feature extraction or the reconstruction of XCT data. We, thus, strive for a discussion on how machine learning concepts can be employed in visual computing in materials science, facilitating knowledge transfer between materials and avoiding errors in training.
    \item \textbf{Immersive analytics in materials science:} We target a discussion of how immersive analytics (AR/VR) can substantially help data analysis and exploration in materials science or if it rather hinders it because of novel interaction concepts, complex metrics for complex data and thus steeper learning curves. Furthermore, it needs to be clarified if current visual metaphors are enough or if tailored representations are required.
\end{itemize}  

\vspace{5pt}
\noindent \textbf{Interpretability and decision making:} The complexity of materials systems, as well as the complexity of state-of-the-art visual analytic tools, features the huge potential to compound challenges in understanding results and making decisions. Specifically, advances in machine learning (ML) are quickly allowing for approximations of large-scale simulations, as well as automated analytics and decision-making. However, in many cases, a human-in-the-loop approach will be required to ensure safe and appropriate outcomes. New techniques will be required to facilitate understanding of automated tools, as well as uncertainty arising in these tools. We will discuss three topics in this direction:
\begin{itemize}
    \item \textbf{Interpretability of machine learning methods:} ML is often used for dimension and complexity reduction, or to approximate models that can quickly provide insights into the data. However, ML is only as good as its training. In the materials domain, without appropriate and careful training, a machine may generate models that, e.g., use infeasible or ill-advised parameter settings---issues that may easily be avoided through the intervention by a knowledgeable, human, expert. To this end, these techniques must be accompanied by visual tools to support the interpretability of what the methods are doing, as well as by new interaction techniques.
    \item \textbf{Visual indicators for decision making} will also be required. This includes visualization methods for input parameter spaces and output quantities of interest. Furthermore, understanding the uncertainty within these methods is paramount for decision-making. Without knowledge of where and why a model might fail, any decision made using these tools may be faulty. While the influence of parameters and resulting uncertainties are topics in visual computing in general, there are likely to be specific challenges in the material domain, which should be identified.
    \item \textbf{Knowledge-driven and guided analysis:} We target a discussion on how visual analytics for materials science can benefit from new concepts, such as guidance and knowledge-driven exploration---especially, in conjunction with the integrated visual analysis of ``rich’’ data mentioned above. As the complexity and richness of the data increases, these topics will become indispensable components of visual analytics solutions aiming at actively resolving a knowledge gap encountered by users, during their analytical processes. 
\end{itemize}

\vspace{5pt}
\noindent \textbf{Overarching connection between topics:} The aforementioned discussion directions are not completely detached from each other. For example, the topology-related topics are closely related to the ``rich'' data analytics, and the integration of ML and immersive analytics are linked to the interpretability topics. This connection is intended, as we would like to create a linkage between the different discussion groups and to discuss future challenges of the domain in a more holistic manner.

\clearpage

\section*{Overview of the Meeting}

The NII Shonan Meeting ``Advancing Visual Computing in Materials Science'' took place in Shonan, Japan from May 13th to May 17th, 2024. The seminar was organized by Christoph Heinzl (University of Passau, Germany), Renata Georgia Raidou (TU Wien, Austria), Kristi Potter (NREL, USA), Yuriko Takeshima (University of Tokyo, Japan), Mike Kirby (University of Utah, USA) and Guillermo Requena (DLR; RWTH Aachen University, Germany). We hosted 27 participants (including 2 remote participations) from all over the world and diverse expertise, who discussed various topics bridging the domains of Materials Science and Visual Computing. All participants held lightning talks, while four domain experts  (Guillermo Requena, Mike Kirby, Mahoor Mehdikhani, Daniela Ushizima) gave longer overview talks. All talks were followed by a Q\&A session and panel discussion, while we allotted sufficient time for working group discussions throughout the 5 days of the seminar. The social event included the traditional seminar outing, including visits of the Jomyoji and Hokokuji temples and attending a traditional Japanese Tea ceremony.
\begin{figure}[h!]
\centering
    \includegraphics[width=1\textwidth]{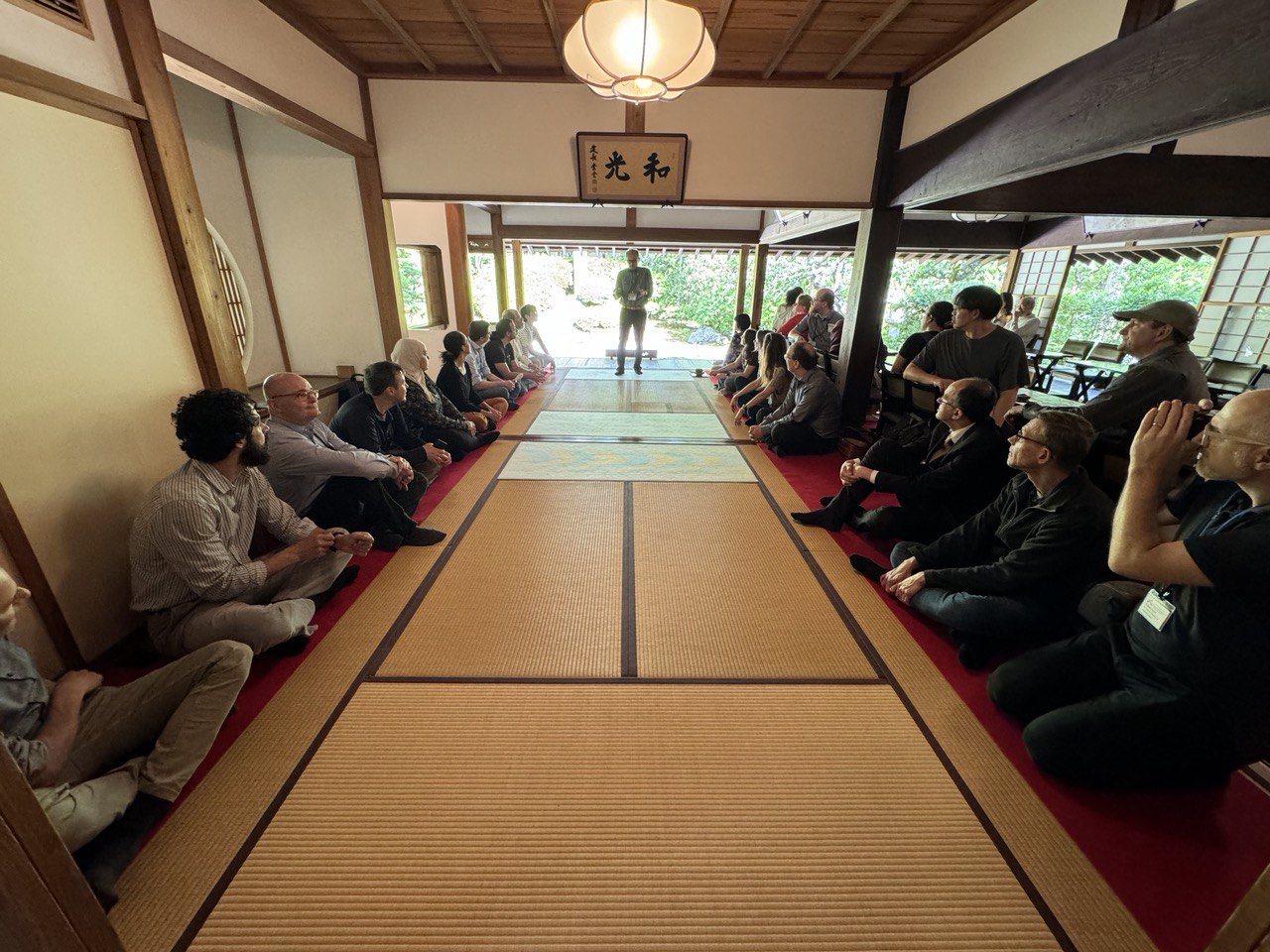}
\end{figure}

\clearpage

\section*{Overview of Talks}
\SHONANabstract{Aerospace materials in the era of the digital transformation}{%
Guillermo Requena\\ 
RWTH Aachen / DLR, Germany}

Traditional approaches to deploy new structural alloys require development cycles that can easily take a decade and are associated with significant economic risks. This is incompatible with current challenges such as global warming, scarcity of raw materials and rising energy costs, which are forcing humankind to rapidly develop sustainable solutions if the level of prosperity of future generations is to be, at least, maintained. To this purpose, innovative and efficient materials solutions along the entire value chain are essential.  In this contribution, examples that highlight how the digital transformation is enabling the acceleration of the design, discovery, development and deployment of new alloys for the aeronautics and space sectors are presented. 
The combination of artificial intelligence, robotic-based labs, high-throughput data generation/analysis, materials combinatorics, predictive simulations, collaborative virtual environments and multi-scale time-resolved experiments result in a suite of data-centric tools whose capabilities will be highlighted through specific use cases: the design of recycled-based alloys, the autonomous testing of metallic structures, the operando study of materials during 3D printing, and the rapid analysis of large experimental datasets. The overall aim is to demonstrate the transformative potential of data-centric approaches to shorten materials development cycles.  

\clearpage

\SHONANabstract{Visualization of Higher Order Tensors in Material Science}{%
Gerik Scheuermann\\
Leipzig University, Germany}

This is joint work with Chiara Hergl (DLR Cologne), Anja Barz, Baldwin Nsonga (both Leipzig University),  Carina Witt, Tobias Kaiser, Andreas Menzel (all three TU Dortmund), Thomas Nagel, Florian Silbereisen (both TUBA Freiberg), Olaf Kolditz (UFZ Leipzig).

The visualization of field data like scalar and vector fields, has been a central part of visual analysis for materials science since its beginning. Also, symmetric tensor fields of second order have seen attention for more than 3 decades now, and the community has also some results on dealing with arbitrary second order tensors. This is different for tensors of higher order than two. However, they play a major role in continuum mechanics and materials science. Properties like stiffness or electromechanical coupling are described by such data, as well as quantities like stress gradients. As modern material science aims at designing and using more and more anisotropic materials, we expect to see more of these fields to be studied in the future. In the talk, we advocate the use of the deviatoric decomposition of arbitrary tensors of any order in three dimensions. This splits the space of all such tensors into the smallest rotation-invariant subspaces and allows to study only totally symmetric, traceless tensors, called deviators. By a construction from Maxwell, one can represent them by a set of directions called multivectors and a scalar, allowing to adapt many well known vector visualization methods for such higher order tensors, In the talk, we concentrate on real examples concerning the stiffness variation in Eulerian coordinates of a biological material represented by an Ogden model, and the variation of electromechanical coupling in a perfect and imperfect lens.
\clearpage

\SHONANabstract{Enabling Visual Analytics of Large Volumetric Data by Error-Bound Compression}{%
Thomas Lang\\ 
Fraunhofer Institute of Integrated Circuits IIS\\
Division Development Center X-ray Technology, Germany}

Modern X-ray imaging strives towards combining the imaging of increasingly larger samples together with finer resolution, i.e., smaller voxel sizes. Consequently, the generated volumetric datasets are getting larger and larger, to the point of classical methods not being capable of processing them anymore. This concerns all tasks in a visual analysis, starting from rendering the 3D datasets up to a detailed analysis of defects, among many more. 

In this talk, we present a volumetric compression scheme exploiting the properties of a three-dimensional discrete wavelet transform in conjunction with quantization and encoding methods~\cite{WaveletCompression}. That method will first express the data in a wavelet basis, for which we choose the Haar wavelet system, followed by the quantization steps and accumulates the result in a HDF5-compatible data format~\cite{SCRWhitepaper}. The properties of wavelets enable a \emph{local} decompression, i.e., the extraction of regions of interest in the original voxel representation without the need for a full decompression. Therefore, it enables the processing of, in principle, arbitrarily large volumes. Furthermore, we give two specific examples of visual analysis of two selected high resolution dataset obtained at a synchrotron facility, one being a 18650 Li-ion battery cell~\cite{CompressionBattery} and a novel matrix metal composite material sample.
\clearpage

\SHONANabstract{Invariants for Materials Science}{%
Hamish Carr\\
University of Leeds, UK}

Mathematicians, Computer Scientists and Domain Scientists all use the term ``invariant'', often with different meanings.  In particular, topological visualization often depends on well-established invariants such as the equivalence of contours or gradient lines, but also on the algorithmic invariants necessary to prove correctness of a given approach.  In contrast, domain scientists may use ``invariant'' to describe a property of a physical system which is known (or expected) to be unchanging.  I will argue that for effective support of materials sciences and engineering, it is necessary not only to deploy existing invariants, but to mine the experience of the materials scientists and engineers to identify invariant properties that can either be mapped to existing techniques, or used as the basis of new techniques.
\clearpage

\SHONANabstract{Statistical Physics in Materials and Data Science}{%
Matthias Sperl, \\
Materials Physics in Space, DLR, Cologne, Germany
\\
Institute for Theoretical Physics, University of Cologne, Germany}

Statistical physics can be seen as being at the foundations of both materials science as well as data science. For materials science, emergent behavior is presented for the special case of granular matter in space where microgravity experiments are investigated for a range of different densities.
\begin{figure}[h]
    \centering
    \includegraphics[height=2.7cm]{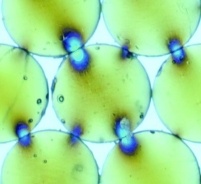}
    \includegraphics[height=2.7cm]{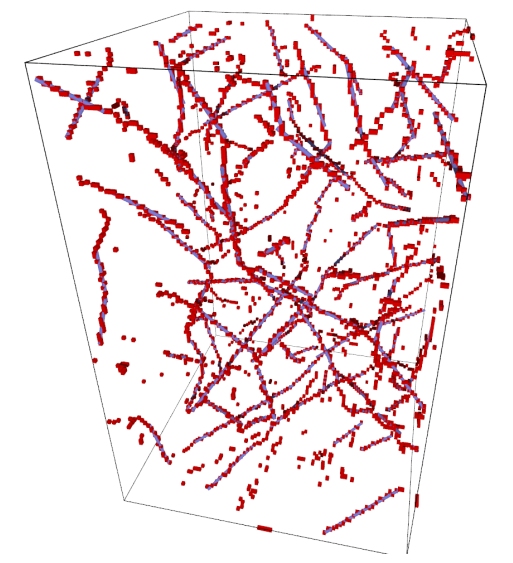}
    \includegraphics[height=2.7cm]{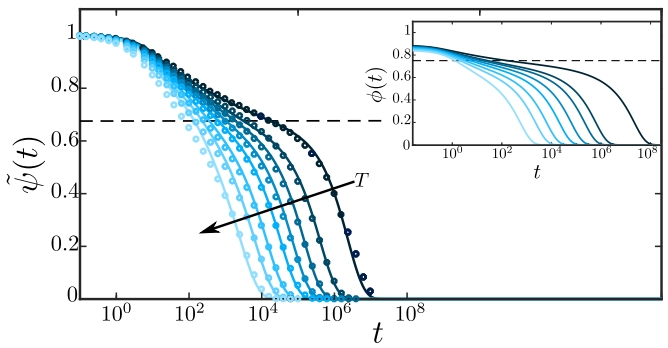}
    \caption{Granular stress birefringence exhibiting forces among individual granular particles (left), adapted from \cite{dlr141788}; dynamics of a granular gas from \cite{dlr135102} (center), and the dynamics of the Fredrickson-Andersen modell
    \cite{onder2019asymptotically} (right).}
    \label{fig:Sperl}
\end{figure}

First on the data science side, machine learning is applied to the data analysis of granular experiments both on Earth and in space. Here, results for the stress-birefringence inside dense granular samples are presented, cf. Fig.~\ref{fig:Sperl} (left). In addition, applying U-Net for image segmentation in a granular gas in microgravity can be used to extract trajectories of individual particles which perform better than traditional ways of particle tracking, see Fig.~\ref{fig:Sperl} (center).

Second on the data science side and more fundamentally, states and behavior of systems in statistical physics can be utilized to encode challenges in machine learning; the Fredrickson-Andersen model \cite{onder2019asymptotically} is for these aspects demonstrated as a physical reservoir for encoding a standard data set like the MNIST data.

\clearpage

\SHONANabstract{Co-design of Materials with Tools from AI/ML and Visual Computing}{%
Mike Kirby\\
Kahlert School of Computing and the Scientific Computing \& Imaging (SCI) Institute, University of Utah, USA}

Depending on your vantage point, the past fifty years could be considered the ``age of computing" or equally the ``age of materials". The past few decades have seen the rise of GPUs, simulation science, and AI/ML as well as the rise of drug engineering, 3D printing, and composite materials. A natural question is how these two powerful and societally relevant fields might join forces. One of the challenges of this ``match made in heaven" is that computer science loves abstractions and materials scientists often live in particulars. Computer scientists seek to look across areas and instances for similarities; materials scientists look across similarities to distinctives.

In this talk, we present an exploration into the intersection of computer science and materials science, probing the intriguing possibilities and challenges that arise when these two powerful disciplines converge. As we delve into this ``match made in heaven, we confront the inherent tension between the abstract nature of computer science and the focus on particulars in materials science. While computer scientists seek commonalities across diverse domains, materials scientists scrutinize distinctions within similarities. By bridging these disparate approaches, we aim to uncover synergies that could revolutionize fields ranging from advanced manufacturing to drug discovery. We start with a discussion of how materials science evolved as a field, discuss example materials problems that motivate the talk, and then highlight lessons learned that might help the area of visual computing engage with materials science research.

\clearpage

\SHONANabstract{Dynamic Perspectives: Visualizing Time and Networks for Analytical Insights}{%
Velitchko Filipov\\
TU Wien, Vienna, Austria}

Dynamic networks are structures where the graph's nodes and/or edges can appear or disappear over discrete or continuous intervals of time.
The visualization and analysis of dynamic networks play an essential role in understanding the structural evolution of a network. The main goals are to support a better overview of the network's evolution and to identify patterns or behaviors. Dynamic networks are most commonly represented as node-link diagrams with the temporal dimension being depicted using animated approaches. 

In this talk, I present a number of different and alternative methods for visualizing and interacting with a network and its temporal dimension and discuss the most pressing challenges currently faced. Furthermore, possible intersections with the Materials Sciences domain are discussed to understand how dynamic network visualization can complement the current state-of-the-art in this domain, provide different perspectives, and facilitate new insights into the data. 
\clearpage

\SHONANabstract{Parameter settings for topology-accentuated visualization}{%
Yuriko Takeshima\\
Tokyo University of Technology, Japan}

The visualization parameter values are critical because they significantly influence the knowledge acquired from the visualization results. The most common approach is to determine these values by trial and error. However, such an approach does not guarantee appropriate visualization results. To address this problem, we proposed a topology-based scheme for setting the visualization parameter values. 

In this talk, I introduce how to set the appropriate transfer function, viewpoint, and illumination position for volume rendering based on the topological structure.
In transfer function design, visualization results can be obtained where the topology structure is easily understood by assigning more colors and high opacities to field values where the topology changes. In addition, new entropy values are defined to determine the appropriate viewpoint and illumination positions, respectively.

\clearpage

\SHONANabstract{Examples from visual analysis in material sciences, granular material and others}{%
Ingrid Hotz, Linköping University}
\begin{figure}[h!]
    \centering
    \includegraphics[width=0.95\linewidth]{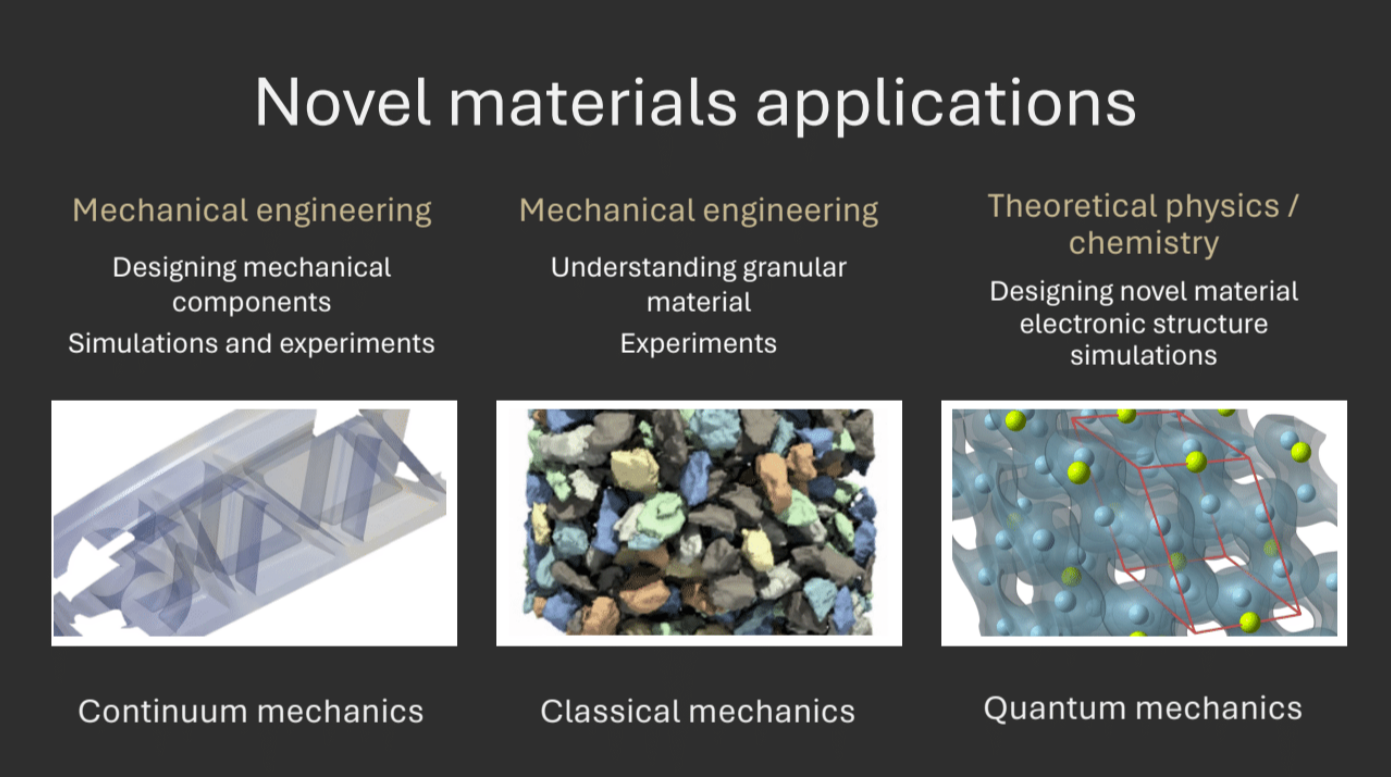}
    \caption{Examples from visual analysis in material sciences, granular material and others - Graphical Abstract}
    \label{fig:Hotz}
\end{figure}
\clearpage

\SHONANabstract{RadVolViz: An Information Display-Inspired Transfer Function Editor for Multivariate Volume Visualization}{%
Klaus Mueller\\
Computer Science, Stony Brook University, New York, USA}

In volume visualization transfer functions are widely used for mapping voxel properties to color and opacity. Typically, volume density data are scalars which require simple 1D transfer functions to achieve this mapping. If the volume densities are vectors of three channels, one can straightforwardly map each channel to either red, green or blue, which requires a trivial extension of the 1D transfer function editor. We devise a new method that applies to volume data with more than three channels. These types of data often arise in scientific scanning applications, where the data are separated into spectral bands or chemical elements. Our method expands on prior work in which a multivariate information display, RadViz, was fused with a radial color map, in order to visualize multi-band 2D images. In this work, we extend this joint interface to blended volume rendering. The information display allows users to recognize the presence and value distribution of the multivariate voxels and the joint volume rendering display visualizes their spatial distribution. We design a set of operators and lenses that allow users to interactively control the mapping of the multivariate voxels to opacity and color. This enables users to isolate or emphasize volumetric structures with desired multivariate properties. Furthermore, it turns out that our method also enables more insightful displays even for RGB data. We demonstrate our method with three datasets obtained from spectral electron microscopy, high energy X-ray scanning, and atmospheric science.
\clearpage

\SHONANabstract{Topological Data Analysis for Materials Science: A Hypergraph Perspective}{%
Bei Wang \\University of Utah}
We discuss the potential of modeling shape data from materials science as hypergraphs to encode multiscale features and higher-order interactions, as well as modeling energy landscapes using tools from topological data analysis. 

Hypergraphs capture multi-way relationships in data, and they have seen a number of applications in higher-order network analysis, computer vision, geometry processing, and machine learning. We introduce a hypergraph distance based on the co-optimal transport framework. Such a distance has nice theoretical properties and can be applicable to study the topological evolution of latent spaces in deep learning. This talk is primarily based on a joint work with Samir Chowdhury, Tom Needham, Ethan Semrad and Youjia Zhou (doi: 10.1007/s41468-023-00142-9).

Formally, A \emph{hypergraph} is a pair $(X,Y)$, where $X$ is a set of \emph{nodes} and $Y$ is a set of \emph{hyperedges} (i.e., subsets of $X$).
A \emph{measure hypernetwork} is a quintuple $H = (X,\mu,Y,\nu,\omega)$, where $(X,\mu)$ and $(Y,\nu)$ are Borel-measured Polish spaces (e.g., probability spaces where $X$ is a Polish space and $\mu$ is a Borel probability measure on $X$) and $\omega: X \times Y \to \mathbb{R}$
is a bounded, measurable \emph{hypernetwork function} that captures relations between nodes and hyperedges (e.g., membership relations, $\omega(x,y)=1$ if node $x$ belongs to hyperedge $y$). The collection of all hypernetworks is denoted 
$\mathcal{H}$. 

Given two probability spaces $(X,\mu)$ and $(Y,\nu)$, a \emph{coupling} $\pi$ is a joint probability measure on $X\times Y$ satisfying 
    $\pi(A\times Y) = \mu(A)$ and $\pi(X\times B) = \nu(B)$ for each measurable $A\subseteq X$ and $B\subseteq Y$. 
The collection of couplings between $\mu$ and $\nu$ is denoted $\mathcal{C}(\mu,\nu)$.

Let $H = (X,\mu,Y,\nu,\omega), H' = (X',\mu',Y',\nu',\omega')$ be two measure hypernetworks. The \emph{$p$-th co-optimal transport distortion} for $p \in [1,\infty)$ is the functional 
$$
\mathrm{dis}_p = \mathrm{dis}_{H,H',p}: \mathcal{C}(\mu,\mu') \times \mathcal{C}(\nu,\nu') \to \mathbb{R}
$$
defined by
$$
\mathrm{dis}_p(\pi,\xi) = \left(\int_{X \times X'} \int_{Y \times Y'} \left|\omega(x,y) - \omega'(x',y') \right|^p \xi(dy \times dy') \pi(dx \times dx')\right)^{1/p}, 
$$
where $\pi$ and $\xi$ are couplings between nodes (i.e.,~$\pi \in \mathcal{C}(\mu,\mu')$) and hyperedges (i.e.,~$\xi \in \mathcal{C}(\nu,\nu')$), respectively. 
The \emph{hypernetwork $p$-distance} is defined to be
$$
    d_{\mathcal{H},p}(H,H')= \inf_{\pi \in \mathcal{C}(\mu,\mu')} \inf_{\xi \in \mathcal{C}(\nu,\nu')} \mathrm{dis}_p(\pi,\xi).
$$

We establish fundamental properties of our hypernetwork distance; in particular, we show that it is a pseudometric on the
space of measure hypernetworks. 
Such a distance can be used for hypergraph matching
and comparison, as well as highlighting simplification levels of interest for real-world datasets.

\clearpage

\SHONANabstract{Topological Descriptors for Nanoporous Materials}{%
Dmitriy Morozov\\
Lawrence Berkeley National Laboratory, California, USA}

Machine learning has emerged as an attractive alternative to experiments and simulations for predicting material properties. Usually, such an approach relies on specific domain knowledge for feature design: each learning target requires careful selection of features that an expert recognizes as important for the specific task. The major drawback of this approach is that computation of only a few structural features has been implemented so far, and it is difficult to tell a priori which features are important for a particular application. The latter problem has been empirically observed for predictors of guest uptake in nanoporous materials: local and global porosity features become dominant descriptors at low and high pressures, respectively. We investigate a feature representation of materials using tools from topological data analysis. Specifically, we use persistent homology to describe the geometry of nanoporous materials at various scales. We combine our topological descriptor with traditional structural features and investigate the relative importance of each to the prediction tasks. We demonstrate an application of this feature representation by predicting methane adsorption in zeolites, for pressures in the range of 1-200 bar. Our results not only show a considerable improvement compared to the baseline, but they also highlight that topological features capture information complementary to the structural features: this is especially important for the adsorption at low pressure, a task particularly difficult for the traditional features. Furthermore, by investigation of the importance of individual topological features in the adsorption model, we are able to pinpoint the location of the pores that correlate best to adsorption at different pressure, contributing to our atom-level understanding of structure-property relationships.
\clearpage

\SHONANabstract{Topological Tools for the Visual Analysis of\\ Time-Varying Data Sets}{%
Christoph Garth\\
University of Kaiserslautern-Landau (RPTU), Germany}

Topological abstractions are used across a wide range of applications to facilitate feature extraction and visualization. Based on a robust mathematical formulation, and equipped with built-in structural simplification, they are easily employed to achieve a structural and abstract data representation in a manner that is robust to noise. Consequently, topological techniques have found their way into many visualization pipelines. However, their role if often found "behind the scenes", i.e., they are used to process or preprocess data, and the result is visualized, e.g. a segmentation.

In this talk, I argue that topological abstraction are also quite useful as user interface metaphors in visual analysis and exploration of datasets. While they require users to invest into understanding and using them, they enable quick navigation of the features of a data set. This is especially useful when combined with other (linked) views in a visual analytics system. I illustrate this on several examples from recent work, especially focusing on time-varying data.
\clearpage

\SHONANabstract{Extremum Graph: Scalable Computation, Segmentation, and Fabric Quantification}{%
Vijay Natarajan\\
Indian Institute of Science, Bangalore, India}

Topological descriptors of scalar functions such as the Reeb graph, merge tree, contour tree, and Morse-Smale complex have been widely studied and applied within the fields of visualization, shape analysis, and computer graphics. The extremum graph of a scalar function is a useful abstraction that stores the connectivity between maxima / minima of the function. It is a subset of the well studied Morse-Smale complex, is considerably simpler, and yet intuitive and descriptive enough for various applications. In this talk, I will describe an application of the extremum graph to the study of granular material ensembles, which motivates the need for fast algorithms for computing the descriptor. Next, I will describe an algorithm for scalable computation of the extremum graph. This algorithm utilizes GPU and CPU parallelism, supports large data sizes, and generalizes to higher dimensional data.\\  \url{https://vgl.csa.iisc.ac.in}\\ 
\url{https://www.youtube.com/c/vgliisc}

\clearpage

\SHONANabstract{X-ray computed tomography and image processing tools used to unravel microstructure and mechanics of fiber-reinforced composites}{%
Mahoor Mehdikhani\\
KU Leuven, Department of Materials Engineering, Belgium}

Fiber-reinforced composites are engineering materials that provide high mechanical properties and great weight-saving capabilities. Being composed of fibers, embedded in a matrix, composites pose heterogeneity in their structure at different length scales, from micro to macro (see figure). Recently, to characterize their heterogenous structure in static or loaded conditions, X-ray computed tomography has been increasingly exploited. This requires technological developments in both image acquisition and image processing. In this presentation, we present methodologies developed for in-situ X-ray (synchrotron) tomography of composites, providing 3D images of different loading steps at high resolutions. These 3D volumes include invaluable data on deformation and damage (which often appears as cracks) of composites at the microscale. However, extraction of this data is not as straightforward. Hence, as the complementary part of the study, image processing tools have been developed/employed. This includes (machine-learning) segmentation of fibers or voids, fiber orientation analysis, detection of various damage mechanisms, development of image-based computational models, super-resolution, synthesis of realistic microstructure images, etc. Despite these developments, quantification of the acquired data has still a lot of room for improvement in terms of accuracy and efficiency, which is where visual computing can come helpful in the area of mechanics of composite materials.
\begin{figure}[h]
    \centering
    \includegraphics[height=6cm]{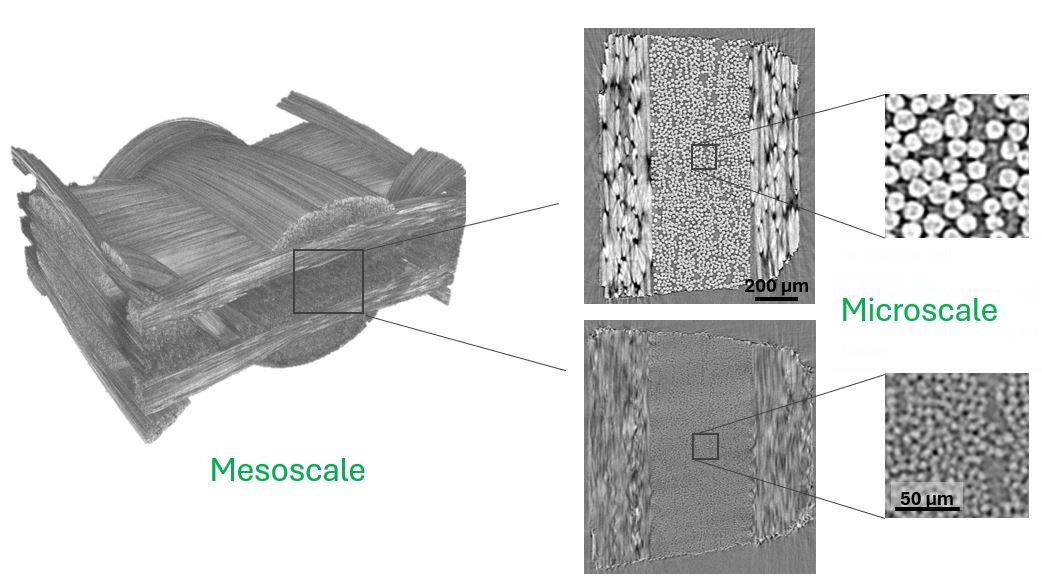}
    \caption{Hierarchy in fiber-reinforced composites, characterized via X-ray computed tomography.}
    \label{fig:mahoor}
\end{figure}\\
\url{www.mtm.kuleuven.be/english/research/scalint/cmg/}
\clearpage

\SHONANabstract{Interacting with DR Projections in Material Sciences}{%
Remco Chang\\
Tufts University, USA}

In this talk, I explore techniques for high-dimensional data visualization and interaction such as to material science data, leveraging continuous latent spaces learned from databases of materials (e.g., molecules, alloys, etc.). The first project, \textit{UnProjection}, employs an autoencoder framework to approximate traditional dimensionality reduction (DR) methods such as UMAP, t-SNE, and PCA. This approach not only enables effective dimensionality reduction but also allows for the synthesis of new data points based on specified 2D positions in the embedding space. By enabling the generation of materials with desired properties directly from their latent space representations, UnProjection opens new avenues for material discovery and optimization.

The second project, \textit{DimBridge}, enhances the interactivity and interpretability of DR projections for domain scientists. In DimBridge, users can select a visual pattern by drawing, such as trails of points, directly onto the DR projection. The system then generates a minimal scatterplot matrix using the original high-dimensional data that best explains the drawn visual features. This bidirectional interaction between the latent space and the original data dimensions allows scientists to intuitively explore and understand complex material datasets, facilitating the identification of underlying patterns and relationships. 
\clearpage

\SHONANabstract{Towards Cross Virtuality Analytics in Materials Science}{%
Christoph Heinzl, \\
University of Passau, Germany}

The analysis and exploration of rich XCT data, including large primary data (e.g., volumetric datasets or series thereof) generated X-ray computed tomography (XCT) as well as secondary derived data (e.g., segmentation masks, labeled data, multivariate data), can be cumbersome in traditional desktop based visualization setups. Often domain specialists require an understanding of spatial data in the original data domain, considering abstract data on attributes of interest in various representations and levels of detail. 

In this talk, cross virtuality anlytics (XVA) \cite{frohler_survey_2022} has been addressed as an emerging field in materials science providing novel means for integrative solutions. XVA enables visual analytics to make use of transitional and collaborative interfaces along the reality-virtuality continuum. It targets to seamlessly integrate suitable visual metaphors, across different devices, collaboratively supporting multiple users for solving a joint analysis task. 
In a number of design studies on VR-based \cite{gall_imndt_2021} and AR-based \cite{gall_mdd-glyphs_2024} immersive analytics techniques as well as a system integrating techniques along the reality virtuality continuum \cite{gall2022cross}, this talk makes the claim, that immersive analysis can provide novel insights into rich XCT data, which have not been possible before. As use-case of the presented methodology, the analysis of fiber reinforced composites is addressed.
\clearpage

\SHONANabstract{Visual Analytics for Explainable Deep Learning}{%
Marco Angelini\\
Sapienza University of Rome / Link Campus University of Rome, Italy}

In this talk, I will provide an overview of the main approaches existing in Visual Analytics literature supporting the explanation of deep learning models \cite{larosa2019}. I will illustrate the works following a proposed categorization in five eXplainable Artificial Intelligence (XAI): feature attribution (where the model focuses on input data), Learned features (which features the model components, e.g., neurons, layers, the model has learned), Explanations by example (which are the most similar input data to the studied instances), Counterfactuals (how and how much do I need to change input data to get a different decision from the model), and Model Behavior (how the models respond to different changes made to it during its whole functioning time).
Visual design to support these explanations will be illustrated, highlighting their novelty aspects and fit to the XAI category.\\
A comparative analysis in visual design usage, coverage of explanation category, support to human user workflows (e.g., exploratory analysis, confirmatory analysis, comparative analysis) support will also be presented between the Visual Analytics discipline and Materials Sciences, with a work that has further enhanced the analysis toward this domain with a preliminary analysis of around 20 contributions from the Materials Sciences focusing on the usage of Deep Learning models and XAI techniques. Preliminary findings highlight differences in XAI category distribution, usage of rich visual environments, and no or low presence of integrated workflow support for the human analysts, while also presenting interesting aspects on the use of surrogate models and the identification of usage of specialized Deep learning models including specific features of materials properties which can be useful for expanding the visualization informativeness and analytics support.\\
The identified findings can be useful in supporting further integration of Visual Analytics for Materials Sciences and shaping further research directions at the intersection of both research areas.
Full analysis through a live explorer is available here:
\url{https://aware-diag-sapienza.github.io/VA4XDL/survis/}\\

\clearpage

\SHONANabstract{Exploring Energy Materials with Machine Learning: From Brains to Lithium Metal Batteries to Biofuel}{%
Daniela Ushizima\\
Lawrence Berkeley National Laboratory, University of California San Francisco, and University of California Berkeley}

In this presentation, we explored innovative work using convolutional neural networks (CNNs) across various fields including neuroscience, botany, and material science. We highlighted several collaborative projects that developed advanced techniques in semantic segmentation, allied with scientific instrumentation, particularly synchrotron X-ray imaging, and domain expertise, enhancing the understanding and analysis of complex biological and material systems.

We showcased key projects involving diverse subjects such as multimodal brain imaging, abnormal protein structures, deleterious formations in batteries, and plant roots. Utilizing state-of-the-art technologies, comprehensive databases, and visualization, these projects pushed the boundaries of our current scientific knowledge, moving toward the automation of experiments recorded as images and the creation of self-driving laboratories. Our goal was to demonstrate how the convergence of these disciplines underpinned our ongoing and future initiatives to advance the interface between energy materials and computing sciences, showcasing the transformative potential of machine learning in scientific research.

More info: \url{https://ushizima.com/}

\clearpage

\SHONANabstract{Topics on Multiple Variables and Interpretability at Fujitsu Research Institute}{%
Daisuke Sakurai\\
Fujitsu, Japan}

This talk shows select research topics on multiple variables and interpretablity happening at Fujitsu Research Institute. 
A focus is given for the analysis of multiple scalar fields. 
Scalar fields arise in analysis of various forms.
One common task is to explore relations between isosurfaces from multiple fields, which is challenging due to the multi-diemensional parameter space of isovalues. In this context, I show some recent work on the Reeb product. The Reeb product is a product space of Reeb graphs, consisting of every combination of two contours, each from a different scalar field. The Reeb product can be partitioned by semantic categorization of the contour-contour relation.

\clearpage

\SHONANabstract{Sputtering to Uncertainty: 
An overview of work @ NREL}{%
Kristi Potter\\
National Renewable Energy Laboratory, USA}

In this talk I will give an overview of some materials research that I’m involved with at NREL and then dive into ensemble visualization work including surrogate modeling and uncertainty visualization for decision making. NREL is one of 17 US national laboratories, and we are located outside of Denver, CO. The lab’s mission is to work on electrical efficiency and renewable energy and that includes work that pushes technological advances in basic sciences as well as analysis of markets and techno-economics.  The computer science center is a growing team within the lab, and our work is helping to develop workflows for materials discovery, including data collection, AI, and augmented reality for autonomous lab operation, as well as studies on critical minerals in supply chain analysis. The work that I am currently focusing on is within the field of uncertainty visualization.  Specifically, we are working in ensemble simulations in which multiple instances of computational model are run with variations in initial conditions and input parameters, resulting in a collection of datasets that help mitigate the uncertainty in a system and give a better understanding of the relationships between input and outputs. With the rise of AI technologies, we can now create surrogate models of the ensemble, allowing us to go directly from input parameters to visualizations almost instantaneously, or, using Gaussian Processes, estimating input parameters that will achieve defined output results within an uncertainty bound. These novel technologies helps get a user a quicker way to understand the characteristics of the model, however the challenges that present include how to actually use such a system for decision making.  This includes how to create visualizations that incorporate context on the science as well as how to distinguish between the forward or inverse problem. We still need new research on how to connect our science outputs to people using those outputs for real-world decisions.

\clearpage

\SHONANabstract{Tools for Visual Analysis of Production and Test Data}{%
Tobias Schreck\\
Graz University of Technology, Austria}

Visual Analytics research can play an important role in understanding data from industrial processes, products and production. Many research institutes and initiatives to date consider the visualization and visual exploration of industrial data. We provide an overview about our current work in this area. In \cite{242f15957ce54d06b04b5bde4b99e213} an approach for analysis of data from the production of high-quality aluminum products is presented. Data is captured along the production process from melting, casting, rolling and ultrasound-based quality inspection. Our prototype allows interactive selection of quality data and its correlation to production data, supporting the search for factors influencing the quality. In \cite{10508047} an approach for the exploration and comparison of simulation and test data in electrical motor engineering is presented. Pairs of time series from simulation and measurement data are plotted in the parameter space of torque and motor speed. Detail on demand interaction allows to drill down and identify e.g., characteristics of the test bed operation, and precision or improvement potential of the simulation approaches. Then, in \cite{MANDALA} we report ongoing work in the visual exploration of anomalies in multidimensional time series data. Our approach is based on the well-known scatter plot visualization and allows comparison of normal and anomalous data points from cyclic measurement data. Such and other existing approaches could also be applied in materials science, and we conclude with proposed future work in the area.

\clearpage

\SHONANabstract{Visualization Literacy \& Education}{%
Christina Stoiber\\
St. Pölten University of Applied Sciences, Austria}

Visualization Literacy is ``the ability and skill to read and interpret visually represented data in and to extract information from data visualizations'' (Lee et al., 2017). Given the growing complexity and volume of data, visualization literacy is crucial for professionals in every field to make informed decisions. However, people have to learn how to interpret and construct visualizations to make informed decisions. A recent examination of youth’s and adults’ ability to interpret and construct data visualizations by Börner et al. (2016) indicates that the general public has a low level of visualization literacy. This hinders people from accessing valuable information. In this talk, some state-of-the-art examples of learning environments and didactical methods are presented that have the potential to increase the visualization literacy level of users. Besides, two projects are presented called ``Self-Explanatory Visual Analytics for Data-Driven Insight Discovery'', ``Vis4Schools: Fostering Information Visualization Literacy in Schools'' and Comixplain. Finally, some potential topics and links to the field of materials sciences are drawn. 

\noindent \url{http://seva.fhstp.ac.at/en/results},\\ \url{https://vis4schools.fhstp.ac.at/}, \\
\url{https://fhstp.github.io/comixplain/}

\clearpage

\SHONANabstract{Harnessing Visual Computing in Material Sciences through Insights from Biomedical Visualization}{%
Renata G. Raidou\\
TU Wien, Austria}

In this talk, I draw parallels between the field of biomedical visualization and visual computing for material sciences. I start my talk by discussing three challenging topics that we are working on in my group: data heterogeneity, uncertainty within ensembles of data, and immersive analytics. In the first topic, I show our recent work on designing and developing a visual analytics solution for the exploration of radiogenomic information together with clinical information in a cohort of cancer patients. Also, I discussed our recent work on the visualization of 4D dynamical systems used to represent biological processes. In the second topic, I present how we deal with uncertainty---as in errors or variability. For the former, I discuss our work on the visual analysis of segmentation errors also in correlation with quantitative attention information and reasoning data from think-aloud protocols. For the latter, I showcase our predictive approach to support the visual analysis of anatomical variability for radiotherapy decision support, including our recent work on integrating guidance in clinical visual analytics solutions. Finally, I briefly introduce our future intentions of employing visual analytics for the design of digital twins targeting patients with atrial fibrillation. 

\noindent \url{https://www.renataraidou.com/}

\clearpage

\SHONANabstract{Visualization of Time-Oriented Data}{%
Wolfgang Aigner\\
St. Pölten University of Applied Sciences, Austria}

In my presentation, I explore the complexities and nuances of visualizing time-oriented data and its possible connections to materials science. Time, as a unique dimension, possesses a rich semantic structure that goes beyond a simple linear sequence. This presentation highlights the need to appropriately model this structure to capture the irregularities and cyclic patterns inherent in time. 
Understanding these intricacies is crucial for the accurate visualization for time-oriented data. Examples will be given for why and how these characteristics matter for time arrangement (linear or cyclic), used time primitives (time points or intervals), and numbers of variables (single or multiple). The \href{https://browser.timeviz.net}{\emph{TimeViz Browser}} is introduced as tools to support the selection of suitable visualization techniques.

In materials science, time plays a role in a number of different contexts. Examples are the thermal history of 3D printing data, the time dependence of production processes, material behaviors over time such as crack growth, or the rate of lithiation. All of these pose potentially challenging application areas in the context of visualizing time-oriented data.

The presentation further delves into currently pursued research topics of visualization literacy, audio-visual analytics, situated visualization, and knowledge-assisted visual analytics. Effective visualization requires users to be literate in interpreting visual data. Research in visualization literacy aims to create onboarding and teaching methods for visual analytics tools to help users extract information efficiently and confidently. In audio-visual analytics, combining visual and auditory data representations enhances human data analysis capabilities. This field seeks to bridge the gap between sonification and visualization, leveraging the strengths of both modalities. Situated visualization is an approach that aims to integrate data visualization into physical environments, making digital information accessible and relevant to users' spatial contexts. Knowledge-assisted visual analytics leverages domain experts' knowledge to make analytical reasoning more effective and efficient. By incorporating explicit expert knowledge into the visual analytics process, we can improve the interpretation and decision-making based on complex data. This method has significant applications in fields like clinical gait analysis and industrial manufacturing, where expert insights can substantially enhance data analysis and process optimization. All of these advanced research topics aim to enhance the effectiveness of data-driven technologies and integrate digital information into physical environments.

\clearpage

\SHONANabstract{Visual Comparison of Material Data Distributions}{%
Anja Heim\\
Fraunhofer Institute of Integrated Circuits IIS\\
Division Development Center X-ray Technology, Germany
}

CoSi \cite{heim_cosi_2021} is a comparative visualization framework aimed at aiding materials science experts in analyzing multivariate datasets of internal structures, including fibers and pores. This framework offers a holistic view, with a tabular overview and three detailed visualization techniques, eliminating the need for time-consuming sequential data exploration. Its effectiveness has been demonstrated through two specific usage scenarios and a qualitative user study involving 12 experts, marking a significant advancement in the field.

Furthermore, we presented AccuStripes \cite{heim_accustripes_2024}, a design space that employs position and color to represent univariate distributions with gaps, spikes, or outliers. This approach integrates nine representations, combining three composition strategies (color only, overlay, filled curve) with three binning techniques (uniform, Bayesian Blocks, Jenks’ Natural Breaks). A crowd-sourced experiment demonstrated that the overlay composition strategy was the most accurate and preferred, particularly with adaptive binning techniques. The results demonstrated that AccuStripes outperformed traditional line charts in terms of accuracy for structural estimation and comparison tasks.
\clearpage

\SHONANabstract{Co-Adaptive Guidance for Human-AI Interaction}{%
Mennatallah El-Assady\\ 
ETH Zurich, Switzerland}

\noindent{}Mixed-initiative visual data analysis systems rely on a process of co-adaptation where human and AI agents collaboratively perform data-driven problem-solving and decision-making.  The co-adaptive process describes the dynamic learning and teaching process these agents are engaged in during their interaction in the mixed-initiative system. \\

\noindent{}In this talk, I give an overview of the state-of-the-art in co-adaptive analysis, highlighting co-adaptive guidance in visual analytics.  Structuring the topic further,  I present a recent paper on deriving a structured guidance typology.  To illustrate how such theoretical concepts can be put into practice, I present two interactive approaches for topic model refinement that employ different types of guidance: speculative execution and single-objective agents.  Lastly, I demonstrate the Lotse library as a practical framework for co-adaptive guidance implementation.  \\
\url{https://ivia.ch/}

\clearpage

\bigskip

\clearpage

\section*{List of Participants}

\begin{figure}[h!]
    \centering
    \includegraphics[width=1\linewidth]{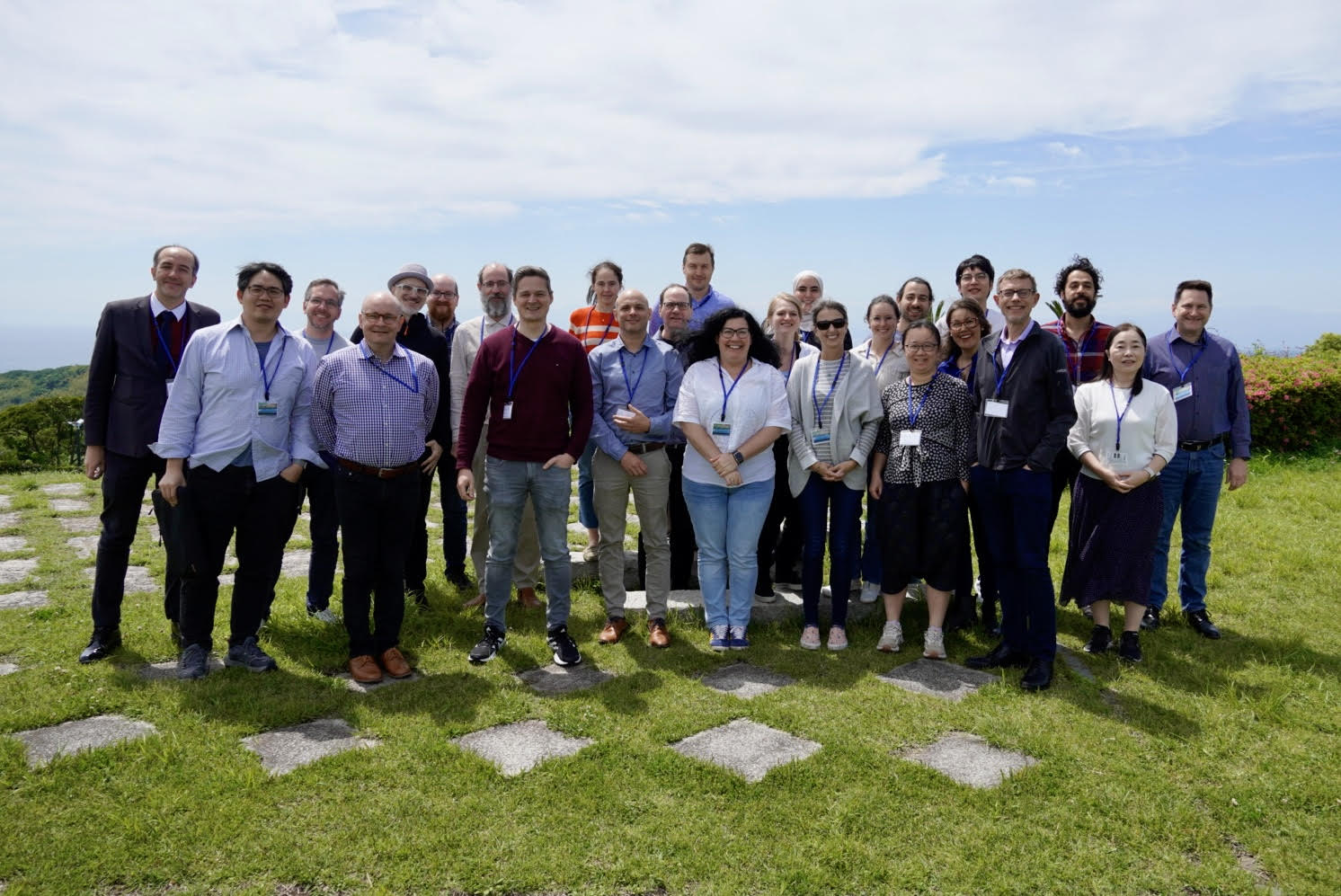}
    \caption{The participants of our meeting.}
    \label{fig:participants}
\end{figure}

\begin{itemize}
    \item Christoph Heinzl, University of Passau, Germany
    \item Renata Georgia Raidou, TU Wien, Austria
    \item Kristi Potter, National Renewable Energy Laboratory (NREL), USA
    \item Yuriko Takeshima, Tokyo University of Technology, Japan
    \item Mike Kirby, Kahlert School of Computing and the Scientific Computing \& Imaging (SCI) Institute, University of Utah, USA
    \item Guillermo Requena, DLR / RWTH Aachen University, Germany
    \item Gerik Scheurmann, Leipzig University, Germany
    \item Thomas Lang, Fraunhofer Institute of Integrated Circuits IIS, Division Development Center X-ray Technology, Germany
    \item Hamish Carr, University of Leeds, UK
    \item Matthias Sperl, Materials Physics in Space, DLR, Cologne, Germany, Institute for Theoretical Physics, University of Cologne, Germany
    \item Velitchko Filipov, TU Wien, Austria
    \item Ingrid Hotz, Link\"{o}ping University, Sweden
    \item Klaus Mueller, Computer Science Department, Stony Brook University, USA [remote participation] 
    \item Bei Wang, Scientific Computing \& Imaging (SCI) Institute, University of Utah, USA
    \item Dmitriy Morozov, Lawrence Berkeley National Laboratory, California, USA
    \item Christoph Garth, University of Kaiserslautern-Landau (RPTU), Germany
    \item Vijay Natarajan, Indian Institute of Science, Bangalore, India [remote participation] 
    \item Mahoor Mehdikhani, KU Leuven, Belgium
    \item Remco K. Chang, Tufts University, USA
    \item Marco Angelini, Sapienza University of Rome / Link Campus University of Rome, Italy
    \item Daniela Ushizima, Lawrence Berkeley National Laboratory, California, USA
    \item Daisuke Sakurai, Fujitsu, Japan
    \item Tobias Schreck, TU Graz, Austria
    \item Christina Stoiber, St. P\"{o}lten University of Applied Sciences, Austria
    \item Wolfgang Aigner, St. P\"{o}lten University of Applied Sciences, Austria
    \item Anja Heim, Fraunhofer Institute of Integrated Circuits IIS, Division Development Center X-ray Technology, Germany
    \item Mennatallah El-Assady, ETH Zurich, Switzerland
\end{itemize}

\clearpage
\thispagestyle{empty}
\begin{landscape}
\section*{Meeting Schedule}
\begin{figure}[h!]
    \centering
    \includegraphics[width=0.8\linewidth]{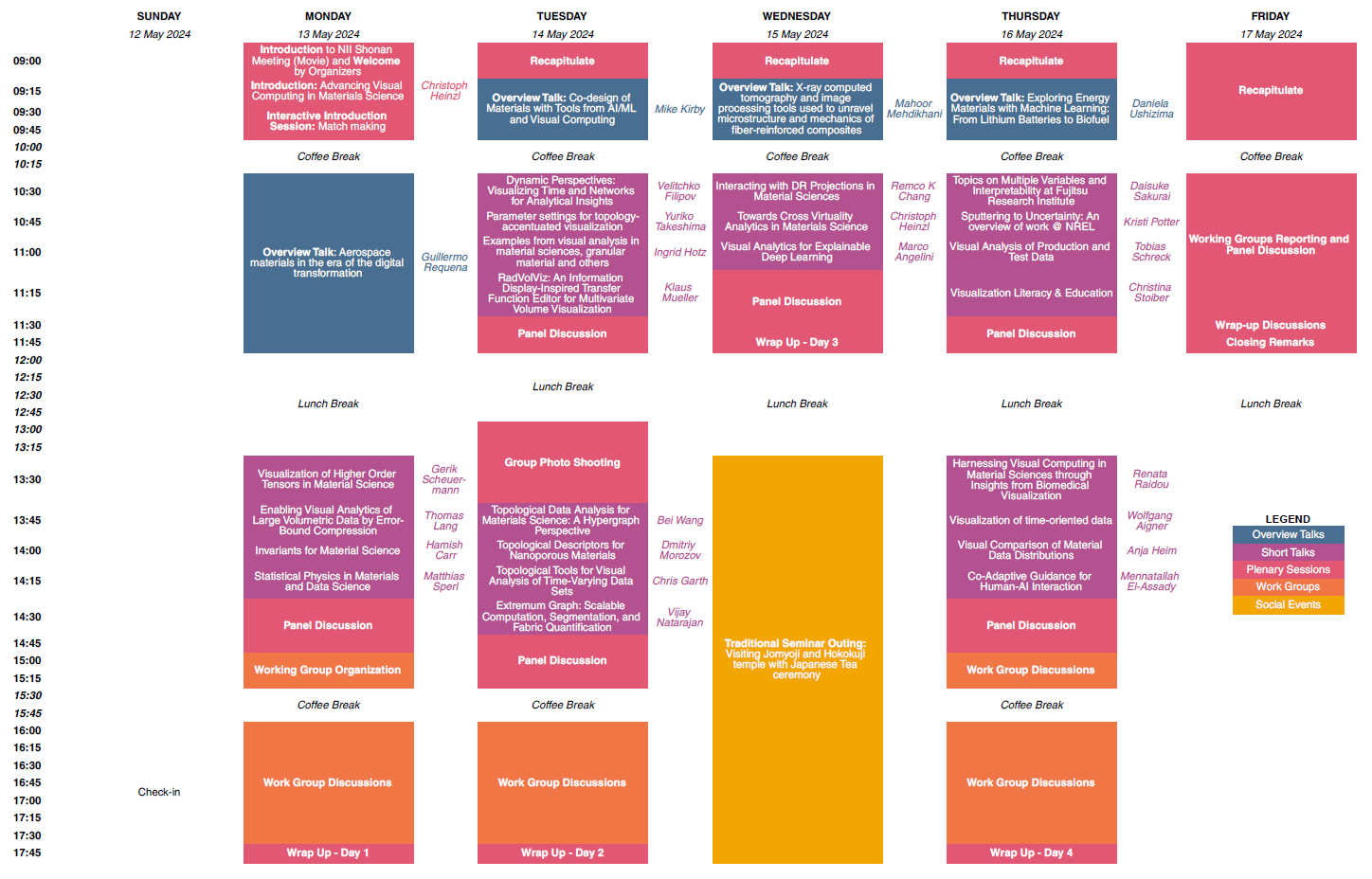}
\end{figure}
\end{landscape}
\clearpage

\section*{Summary of Discussions}
\SHONANabstract{Monday Working Group 1: The Guillermo Group}{%
Daisuke Sakurai, Ingrid Hotz, Remco Chang, Anja Heim, Christina Stoiber, Mennatallah El-Assady, Renata Raidou, Christoph Heinzl}

\begin{figure}
    \centering
    \includegraphics[width=0.9\linewidth]{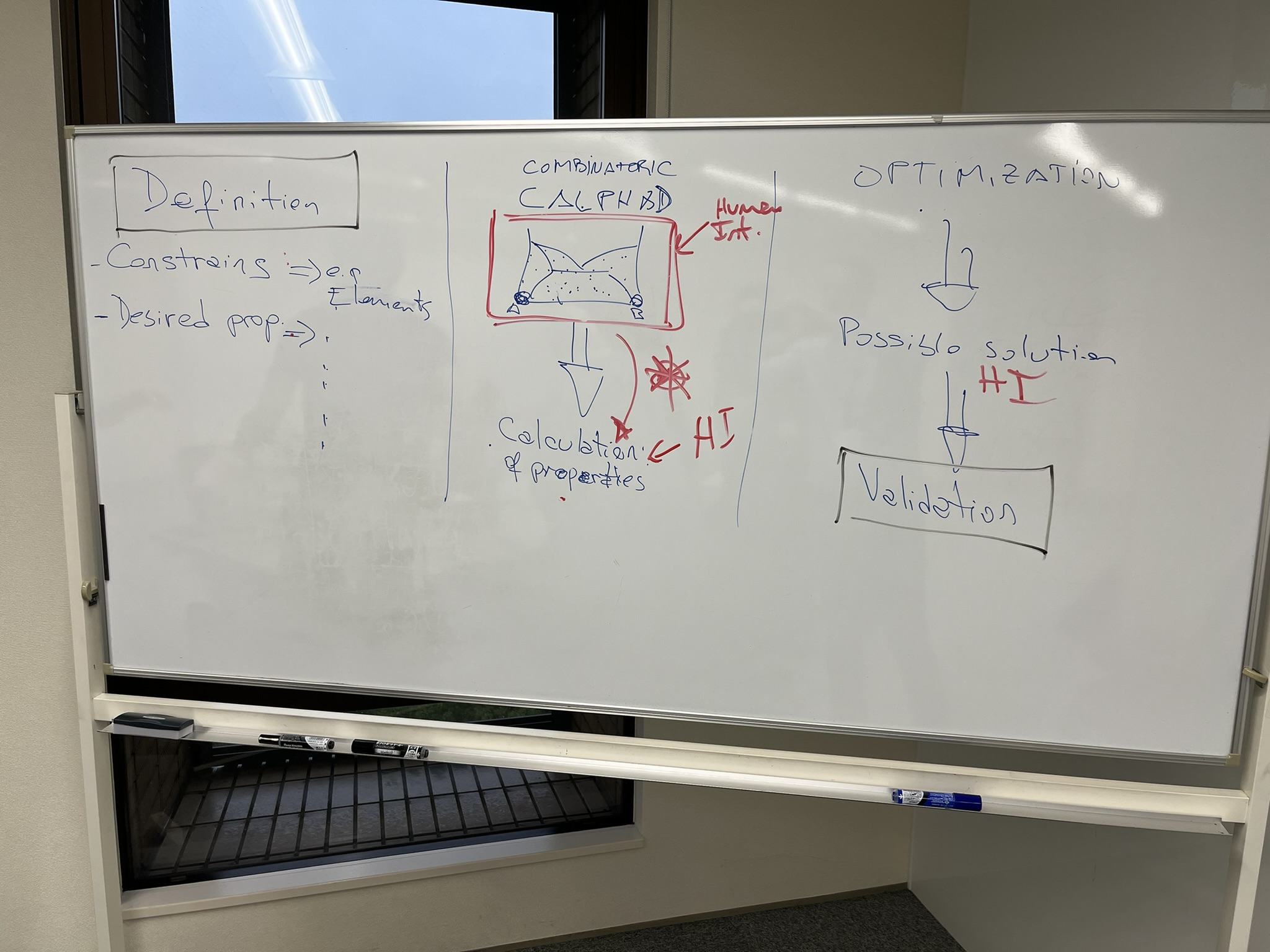}
    \caption{Materials sciences workflow for the discovery of new alloys}
    \label{fig:workflow}
\end{figure}

\begin{figure}
    \centering
    \includegraphics[width=0.9\linewidth]{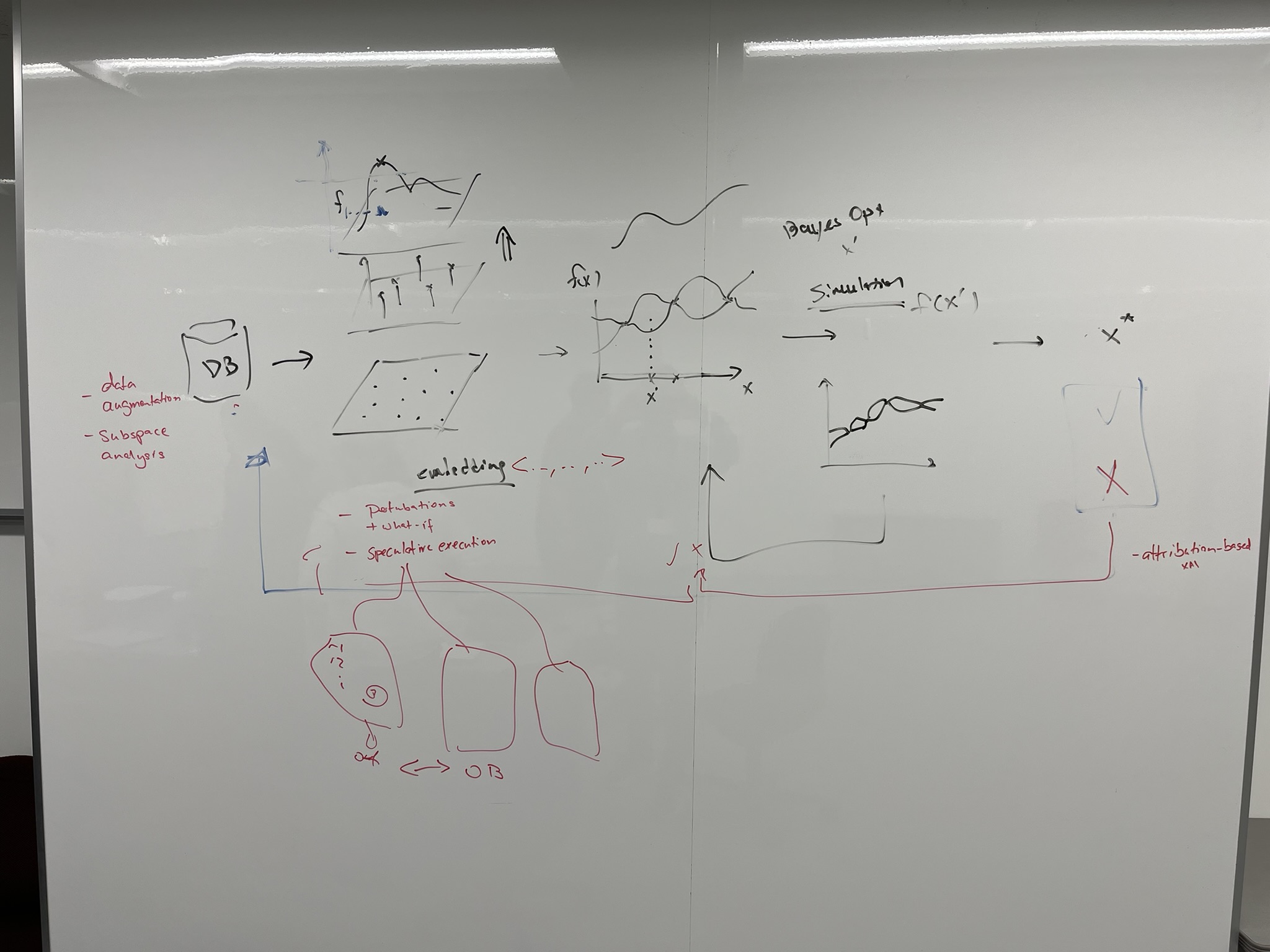}
    \caption{Visualization viewpoint on the material discovery workflow}
    \label{fig:vis_perspective}
\end{figure}

The discussion started with a focus question on "where is the human in the loop in material sciences" and where visualization and visual analytics can contribute. An example could be phase diagrams integrating up to 20 different characteristics of materials: Simulations are used for the exploration of material candidates. Parameter space analysis is required to understand and find suitable candidates to get a direction for research, leaving physics aside. Oversimplification on the physical models lead to candidates which are not suitable, not manufacturable or too costly. This means that conventional optimization techniques (e.g. Bayesian optimization) will not work out. The human in the loop will integrate "physics" by experiments. So, considering a specific material decomposition or starting alloy, first suitable parameter ranges are defined (e.g., wt.\%). Then a material simulation is applied to the input parametrizations in order to compute and simulate material related properties. Currently, input and output parameters are visualized using parallel coordinate plots.The discussion addressed when visualization is required, e.g., for understanding the problem (posthoc), for decision making, etc. Furthermore, the question is, if visualization can contribute beyond that, e.g., visual steering of the data generation. The workflow was described as shown in \autoref{fig:workflow}. A different viewpoint on this workflow is given in \autoref{fig:vis_perspective}.
We figured out that common visual representation in the area of Materials Sciences such as the phase diagram may actually hinder the material discovery process. It boiled down to visualization of the intermediate simulation results (currently triangular representations encoding simulation characteristics against materials) used for setting up hypothesis, a visualization of final results in a higher dimensional representation, as well as using visual steering. Furthermore, alternatives to existing and well established visualizations such as the phase diagram would be thinkable and even desirable for materials science experts. Finally, it took some time to find a common understanding of the problem, goals as well as the used terminology.

Some further interesting questions and topics identified and discussed were the following ones:
\begin{itemize}
    \item Multi-objective optimization + parameter space analysis
    \item Evolutionary modeling?
    \item Interestingness measures?
    \item Simulations are not always representative
    \item What’s the input provided by humans?
    \item Who are the stakeholders?
    \item When are experiments run? How costly?
    \item Provenance tracking?
    \item Wicked problem?  (parameter search space)
    \item Quality metrics vs. optimization functions (simulations) vs. experimental results?
    \item Phase diagrams → Models to calculate properties
    \item Decision-boundaries: Perturbation analysis
    \item Educated guesses: speculative executions? (causal non-linear relations)
    \item Sensitivity analysis
    \item Uncertainty modeling?
    \item Human feedback integration?
\end{itemize}

\clearpage

\SHONANabstract{Monday Working Group 2: The Mahoor Group}{%
Mahoor Mehdikhani, Kristi Potter, Gerik Scheuermann,
Thomas Lang, Dani Ushizima,	Marco Angelini,
Velitchko Filipov	
}

The goal of our session was to better understand Mahoor's science, specifically fiber reinforced composites. The following items have been discussed:

\begin{itemize}
    \item \textbf{X-ray Computed Tomography (XCT) for Reinforced Materials:}
    \begin{itemize}
      \item Utilization of XCT to study both tabletop and synchrotron setups.
      \item Need to segment internal structures in materials such as carbon or glass fiber reinforced polymers/composites.
    \end{itemize}
    
    \item \textbf{Specifics for Material Types:}
    \begin{itemize}
      \item Analyze carbon fibers with a minimum diameter of 7 microns.
      \item Conduct in-situ testing to observe material behaviors under applied stress directly.
    \end{itemize}
    
    \item \textbf{Segmentation Techniques for Damage Analysis:}
    \begin{itemize}
      \item Focus on the segmentation of cracks and damage within the material structure.
      \item Use advanced methods such as convolutional 2D segmentation for a detailed breakdown.
      \item Train the segmentation model on several slices to improve accuracy in detecting voids and cracks.
    \end{itemize}
    
    \item \textbf{Challenges in Segmentation:}
    \begin{itemize}
      \item Differentiate between voids and cracks, crucial for accurate damage quantification.
      \item Include priors to manage very low contrast issues, particularly relevant in glass-epoxy and carbon-epoxy composites.
    \end{itemize}
    
    \item \textbf{Finite Element Model Development:}
    \begin{itemize}
      \item Create finite element models that simulate material behavior under various conditions.
      \item Reflect real data characteristics, noting the typical presence of nearly two bundles per ply in the material structure.
    \end{itemize}
    
    \item \textbf{Material Failure Analysis:}
    \begin{itemize}
      \item Analyze material failure due to strain to understand critical stress points and thresholds.
      \item How can topological analysis help? 
    \end{itemize}

\end{itemize}

\clearpage

\SHONANabstract{Monday Working Group 3: The Matthias Group}{%
Christoph Garth, Mike Kirby, Hamish Carr, Matthias Sperl, Yuriko Takeshima, Wolfgang Aigner, Tobias Schreck, Bei Wang, and Dmitriy Morozov}

We had planned to start with a discussion of use cases, challenges, and pain points in materials science. However, we went over the three areas/partitions of the visualization community: topology-based tools, integrated visual analysis, and interpretability and decision-making.

A typical use case would be to produce the best heat-resistant material for re-entry vehicles.  We then moved to discussing visualization as a way of exploration versus explanation. This discussion transitioned to energy landscapes. Matthias has an entire host of gradient-based methods. The challenge is that the coordinates used for simple visualizations are often not the best for the actual problem at hand. The question of ``what are you trying to do"? The answer is often to find the global minimum or at least a good minimum. There were particular questions to understand what the process of optimization realizes. We then discussed how physical cooling might connect to physical cooling, and then beyond that, what the landscape looks like. 

Visualize the entire landscape, or the path within a landscape, or compare landscapes. Many of the questions then revolved around understanding the energy landscape, what is the dimension, what the gradient and Hessian mean in that context, etc. Discussion of a glassy landscape versus a traditional ML landscape.  Open question: can I characterize with limited information which landscape I have? 
One topic we did not discuss explicitly is uncertainty visualization. 

\clearpage
\SHONANabstract{Tuesday Working Group 1: The Guillermo Group}{%
Daisuke Sakurai, Ingrid Hotz, Remco Chang, Anja Heim, Christina Stoiber, Mennatallah El-Assady, Renata Raidou, Christoph Heinzl}

The group figured out that following our discussions we are coming to a stage where we are becoming quite detailed, which means that we need different people with detailed knowledge in specific areas to discuss.
The group suggested starting the discussion the other way round: vis people explained what they are doing and how this could fit in.
In our group, we had visualization specialists active in the following areas:

\begin{itemize}
    \item Guidance,
    \item provenance,
    \item visualization design,
    \item uncertainty,
    \item scalability,
    \item from proof of concept to real application,
    \item exploring high dimensional data spaces,
    \item plotting energy landscapes,
    \item interaction with projection spaces,
    \item visual analysis of rich XCT data.
\end{itemize}
 
Currently, Avizo is used most by our domain specialist, but in addition Paraview, Python libraries, ImageJ.
As a next step real data is required as well as concrete tasks. Furthermore, funding is required.
As concrete quickly starting topic data can be provided as a use case for existing research.
A series of talks should be organized after our Shonan seminar on respective topical areas to dig deeper into areas and problems.
A potential call for joint proposal on European side could be the Oscar Call: \url{https://oscars-project.eu/}.
A considerable amount of time also went into a discussion on 1D embeddings of higher dimensional parameter spaces (up to 50 dimensions).
\clearpage
\SHONANabstract{Tuesday Working Group 2: The Mahoor Group}{%
Mahoor Mehdikhani, Kristi Potter, Gerik Scheuermann,
Thomas Lang, Dani Ushizima, Marco Angelini, Velitchko Filipov}

In our working group discussions, we collaboratively explored both established and emerging methodologies for quantifying damage in carbon and glass fiber reinforced polymers/composites using X-ray Computed Tomography (XCT). Utilizing both tabletop and synchrotron setups, Mahoor has acquired multiple datasets and conducted diverse analyses using tools such as Insegt and Avizo. Our group exchanged various perspectives on advanced image segmentation techniques, including convolutional 2D and 3D segmentation, to identify fibers and analyze structural damages such as cracks. A significant advancement from our discussions was understanding the array of strategies that Mahoor explored, which included structure tensor analysis—a technique that enhances our understanding of the material's microstructural orientation and integrity. Furthermore, we deliberated on how incorporating the segmentation of individual fibers and utilizing a Deep Convolutional Generative Adversarial Network (DC GAN) to create synthetic datasets could enhance the diversity and quality of our training data. Our group also focused on potentially integrating priors to improve visibility in low-contrast conditions and elaborated on creating finite element models that closely replicate real material behaviors under strain. Further discussions also included methods for visualizing fiber bundles and distance between yarns.

\clearpage
\SHONANabstract{Tuesday Working Group 3: The Matthias Group}{%
Christoph Garth, Mike Kirby, Hamish Carr, Matthias Sperl, Yuriko Takeshima, Wolfgang Aigner, Tobias Schreck, Bei Wang, and Dmitriy Morozov}

Granular materials seem to be a nice training ground for testing various tools. Changing grain sizes, shapes and distributions can help us engage with tools (and test their strengths and weaknesses). The topological tools can be used when dealing with simulation results.  The nice thing with simulation results is that you can compute two-point and multi-point (two-body and multi-body) interactions. In experimental results, it seems to that people are limited to two-point correlations. The question is whether this limitation is on the analysis side, or on the data acquisition side. It seems that it is the latter: the primary limitation seems to be how data is acquired. 

You need to distinguish between three-body forces and three-body interactions. There as a shower of different ways to examine multi-body interactions. Two-body interactions can bootstrap to multi-body correlations. We then started discussing various possible ways of visualizing multi-body correlations.

Structural correlations are of principle interest to material scientists.  Moving beyond two-body to all the possible interaction functions would be of interest. Visualizing those in various ways would also be of interest. For the higher correlations, there are two questions: what happens with large three-body terms.  Where is what we need in the end within the periodic table.  To summarize: in order to understand higher-order interactions: you can use granular systems to study perturbation of the system by making things very large, and try to understand how large things need to be so that you get the behavior you are interested in studying. Try to understand what higher-order interactions and how to draw them? 

The key task is to create a visualization of the triplet configuration. Example: SiO2 and then build up. 

What are the properties of these systems that are of interest? For instance, predict the shear modulus and the young's modulus. 

Question to Matthias: can you come up with a test that would show whether the visualization comes up with a false negative? 

The materials scientists have as a pain-point that they have a failure of imagination to understand what the three-body interactions/correlations look like, and upon that get an intuition on how they behave. 

Who would be the users of such a visualization tool?  A single user, teams of users, interaction, etc.? You would start by looking at how the term impacts the amorphous and/or crystaline structure. SiO2 is relatively simple. SiO2 like structures would be helpful.

We then discussed if this is a one-use tool, or a tool that gets integrated into the workflow of all materials scientists. 

\clearpage
\SHONANabstract{Thursday Working Group: Wrap-Up Discussion}{%
All remaining participants as of Thursday}

Is there a grand challenge project that we can formulate?  Such a question normally starts by picking data and tasks. Is there data available on which we can focus? Maybe we should pick some material system from which we figure out the data and the tasks.

To get some of these methods running, we need to drown these methods in data.  

Based upon things that we have seen in machine learning, 'good' data is better than more data.  What does 'good' mean?  Descriptive data, etc., is important, not just the volume of data. It was brought up that we should maybe start with the point of ``what data do we naturally have (as materials scientists?" and start from there.  We should not generate data that makes things easy. Instead, we should find data sets that are just as we would get in real materials science problems and use those.

Typically, in previous challenges, there was some data for which you had some idea of the things for which you were searching (i.e, there are some gems in the data to find), and that your methods should 'at least' find the things you know to be there.  In addition, it is interesting to find new insights that might come from the data. It is of more interest when there are possible non-intuitive relations (and/or correlations) that are not naively apparent in the data.  The new analysis and visualization methods will hopefully find and/or establish where he/she should possibly look next in terms of data exploration.  It is good to have a test in which there are certain relations that are there for which the visualization scientists are not told it is there, as this helps to make sure that there are no false positives.  Such an plan creates a training ground to help engage the visualization researchers.

What is the next generation of Ashby diagram?  Consider a tool that the materials scientist would use to get good starting points of what are the hidden relations in the data (given certain known relations in the data as a starting point). 

On a different note, we need infrastructure that will allow us to even run these challenges.  How do we facilitate a stable infrastructure that can be used cross groups, areas, and data on these types of challenges?  We have not had a way to create a sustainable track within the visualization community. 

Note that Ashby diagrams are used in many but not all areas of materials science. A different angle is to study the performance of a material relative to its microstructure (e.g., grains, etc.).  This is a another (hidden) issue that would be of interest. These questions are more related to the processibility of the material.  

Maybe we should consider the entire pipeline from the microstructure to the characterization to the high-dimensional Ashby diagram.  At the same time, this pipeline view of a challenge will need a software framework that would enable different people to engage in different places within this pipeline.

A question of how to we encode the ``data hunches" of the materials scientists into our visualizations. 

\begin{figure}
    \centering
    \includegraphics[angle=-90,origin=c,width=0.75\linewidth]{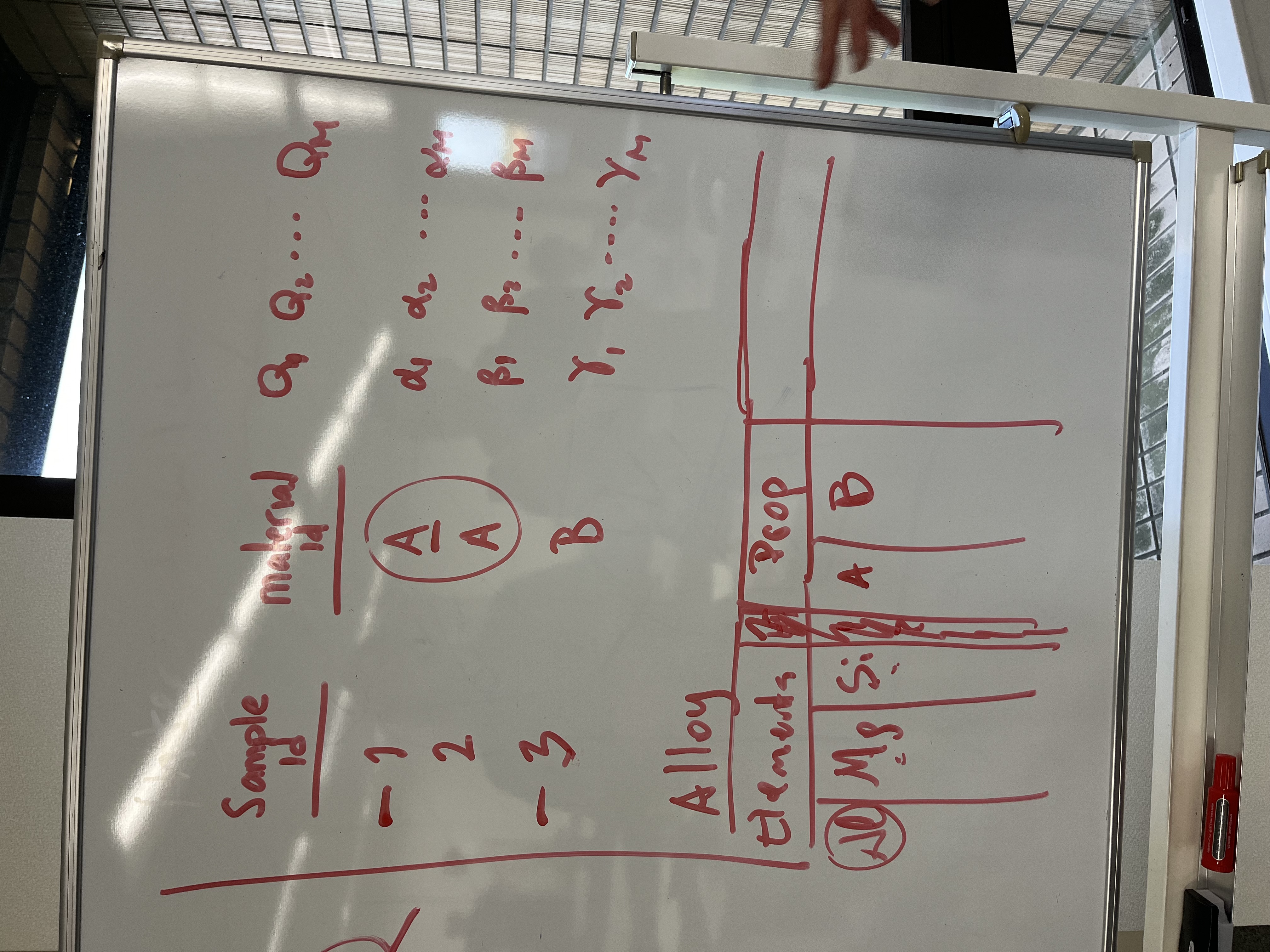}
    \caption{First description of the data visualized in the Ashby diagram}
    \label{fig:ashby_data}
\end{figure}

A recommendation was that we work through the workflow of the materials scientists -- for instance, how are the dots on the Ashby diagram generated, or the pipeline from material to image to segmentation to meshing to simulation to QoI -- and see if we can challenge each of the meta-boxes and come up with abstractions on which individual researchers and/or groups in the visualization community can work. A first characterization of the data is given in \autoref{fig:ashby_data}.

\clearpage

\SHONANabstract{Panel Discussion 1 (Monday, 13 May 2024, afternoon session)}{%
}
This panel consisted of the following four speakers: Gerik Scheuermann, Thomas Lang, Hamish Carr, and Matthias Sperl.  Gerik started us off with a discussion of tensor decomposition and visualization methods. He was followed by Thomas, who brought us to the world of image processing used for materials science, with a focus on enabling visual analytics of large volume data through error-bound compression. Hamish helped the audience consider invariants for materials science. Lastly, Matthias exposed us to granular materials and their use in aerospace applications.

Matthias asked a question of Gerik concerning $C_{2222}$. Gerik did not recollect anything in particular that his analysis showed. They agreed to follow up offline. 

The next question, from the audience, was concerning Thomas' compression and decompression. There was a discussion concerning how it works and the worst-case situations.  This led to a discussion of the global/local nature of the decomposition. 

Matthias was then asked about granular viscosity.  It is close to that of a fluid; however, unlike in a fluid, the relationship between rate-of-strain and stress are different.  When things are acting like a fluid, you obtain something like a viscosity.  As the material hardens, you get different behavior. There was then a discussion of how you infer things from the data.

For Hamish, there was a question concerning the crystal isometry. The question was whether it was stored in as a network. There was a discussion concerning that the files contain and how you infer the invariants. The question from the audience was whether there were other motif-type analysis tools that could be used for studying invariants.

To Gerik an audience member asked the following question. You decomposed to symmetric parts and non-symmetric parts.  The more complex the visualization, the less likely the materials scientist to use it was the assertion.  What is your experience? Gerik related that there was limited interested by some material scientists.  However, on the technical analysis and modeling side, they were very interested in getting any visualization methods that could be used for high-order tensors. Hamish, in the conversation, introduced the idea of partitioning the space as a way to help create new, more simple visualizations.

For Thomas, is thresholding the ``secret sauce" or something else. The answer was a discussion concerning the general nature of the reconstruction. It was acknowledged that the process is meant to be general purpose for everyone doing compression of beam-line data.  It is assumed that almost everyone wants to get rid of noise.  Note that there are loss-less variants.

For Matthias: what are the pain points that can be discussed in the materials workshop discussions. Matthias said that we have one point of grounding: we want to respect the physics.  The question then is how much you are willing to 'violate' the properties that you believe you know to give you a flexibility to explore. 

What is the role of human interaction in what you (all) are presenting? It looks like the selection of parameters would be one choice.  Is there any other place for stakeholders to interact? Each panelist talked about how parameter variation modified through user interaction would impact their work.

The last question was a ``technical" (detailed) question concerning Matthias' U-Net statement. Matthias answered with walking the audience through his process. What they learned was that with the U-Net they could trace particles much longer than previous thought.

\clearpage
\SHONANabstract{Panel Discussion 2 (Tuesday, 14 May 2024, morning session)}{Velitchko Filipov, Yuriko Takeshima, Ingrid Hotz, and Klaus Mueller (remote)
}
This panel consisted of Velitchko Filipov, Yuriko Takeshima, Ingrid Hotz, and Klaus Mueller (remote). Velitchko started the session by talking about dynamic networks and their role in visual analytics. Yuriko followed with a talk on parameter setting for topology-accentuated visualization. Ingrid then presented examples from visual analysis taken from materials sciences collaborations. Lastly, Klaus presented his tool RadVolViz, an information display-inspired transfer function editor for multivariate volume visualization.

The first question was for Yuriko concerning what scalar she was visualizing.  She answered that it was the pressure. There was then a general audience question to all members: there was a discussion of domain-specific diagrams.  Based upon this, how do you span the divide between their specific diagrams and what we do.  Ingrid answered that although we have many cool tools, it is important to acknowledge that if we do not connect it with something that they know and/or understand, it will be difficult for them to fully engage.  We need to be careful not to overwhelm our colleagues, but instead select small steps that aid them in engaging. Ingrid then gave an example from her experience working with chemists, in which the temptation was to try all the flashy things, but starting with the naive things allowed them to build respect with their collaborators and then aid them to be courageous. The audience asked how to know when to be courageous?  The answer by Ingrid is that part of it is building trust with the collaborator over a long period of time working together.  In doing so, building long-lasting trust relationships, in which you invest a lot, are needed. To Ingrid, all problems are interesting, so select collaborations based upon the people. Velitchko echoed the same sentiment; however, there are some steps that you can take.  Velitchko often tries collaboration by immersion, which helps engage the collaborators in a hands-on way. Ingrid added that probably the most important thing is mutual respect -- appreciating how much work went into the generation of the data. Klaus emphasized that you need to learn their language, read their journals, and spend the time to read about the data sets and why they matter. He would then demo them something, and invariably they would want something different.  You really need to meet with them all the time, at odd times and when they have time. 

A technical question for Klaus: you have an option to change the sequence and consequently change the colors. Klaus points out that they cannot flip the colors in this particular tool.

A question then to all: the topological methods appear to have huge potential to handle data with high dimensions. However, it is an abstraction of real data. Materials scientists like to have 'real abstractions' (connections with the physics, etc.).  Are we losing the intuitiveness? Ingrid answered by confirming that they first thing they did was three-dimensional rendering of the data, which allowed them to get the intuition.  From this they could then come up with different visual abstractions and allowing the materials scientists to connect the new visual metaphors with their physical intuitions. Ingrid made clear that you have to be very careful in your choice of abstractions and how you explain them to your collaborator.  Once you convince and accept them of the value, the materials scientists then need to be willing to help transition these tools to their fellow materials scientists. Velitchko reiterated that the goal is not to generate things that are non-intuitive.  In his work, they often do collaborative design so that both groups get onboarded together (through mutual interaction with the data and the visualization).

Klaus mentioned prototyping and trying, prototyping and trying. Velitchko commented that you often have a toolbox and re-assemble things.  So there are cycles of 2-3 months for design, prototype, and try deploying to the users.  

Ingrid was asked about the type of analysis that was done on the fiber-reinforced material.  What type of problems were you addressing?  Ingrid pointed out that they work she showed was to demonstrate to the industrial community the capabilities of visualization.  She did not know the particular question and company that was using the data, but her collaborator wanted her to highlight the power of these new visualization methods.  For her, the opportunity to be an ambassador from the visualization community to the industrial community was the key.

\clearpage

\SHONANabstract{Panel Discussion 3 (Tuesday, 14 May 2024, afternoon session)}{Bei Wang, Dmitriy Morozov, Chris Garth, and Vijay Natarajan (remote)
}

This panel consisted of the following four speakers: Bei Wang, Dmitriy Morozov, Chris Garth, and Vijay Natarajan (remote). Bei started us off with a talk on topological data analysis of materials science via a hypergraph perspective. Dmitriy then presented topological descriptors of nanoporous materials. Thirdly, Christoph presented a talk on topological tools for visual analysis of time--varying data sets. Lastly, Vijay presented (remotely) a talk on extremum graphs as a way of doing scalable computation, segmentation, and fabric quantification.

The first question was addressed to Bei: about optimal transport, you said it was easy to compute. The audience member said they disagreed.  He wanted further details on the nature of the claim.  Bei explained her statements and how the data she showed scaled reasonably. She also acknowledged that she used highly-optimized libraries for all the computing. Dmitriy wanted to dig into why the audience member questioned (and begged to differ).  The audience member said that in his mind, 1d items could be done quickly, but things are more complex in multiple dimensions. 

Another question: why is the Morse-Smale flow better suited than watershed?  The question was addressed to Vijay.  Theoretically, the MS and watershed have similar properties. One of the benefits of the MS is the ability to store both the segmentation and the compact network.  MS allowed for an ability to handle noise (in over and under segmentation) which gave it an advantage over the watershed method. As a follow-up question, if you used a collection of marbles, then watershed would beat MS? Vijay answered that for nice convex shapes, both methods would do well (and similarly). There was then a detailed question concerning Vijay's contact networks slides.

A question for Bei: when we try to extract properties from structural features, we often use two-particle interactions and two-body correlations, three-body, etc.  When we do that, you immediately get into the domain of hypergraphs. The hypergraphs encode not only pairwise relations, but three-way (and more) relations.  The talk by Bei introduced a way of connecting the higher-order correlations (not just the pairwise correlations).  There was then a question of whether there are more intuitive ways to model and portray higher-order relations. There was then a general discussion of the value of drawing hypergraphs.

A general question was about the boundary to entry for using the techniques that were presented. Christoph said that for what he showed, the software is well known and used (TTK).  For Dmitriy's work, he mentioned that you can always create graphs without domain insight.  However, how do you bake physics into the representations and descriptions? For making sure that the geometry and chemistry are simultaneously respected takes hand-tuning. 
\clearpage
\SHONANabstract{Panel Discussion 4 (Wednesday, 15 May 2024, morning session)}{Remco Chang, Christoph Heinzl, and Marco Angelini}

This panel consisted of the following three speakers: Remco Chang, Christoph Heinzl, and Marco Angelini.
Remco started the session by presenting interactive dimensionality reduction (DR) applied to materials.  Christoph then presented cross virtuality analytics in materials science. Lastly, Marco presented visual analytics for explainable deep learning.  

There was a talk from the audience concerning Remo's talk -- in particular, asking about how to possibly dig into the high-dimensional space (e.g., 50 or more dimensions).  It was suggested that there might be topological data analysis (TDA) techniques that might help. 

To Marco, surrogate models and deep tree approaches typically come up in the XAI literature. Where do they show up in what was presented? Marco highlighted that he is using a referenced categorization (not his own). 

There was then a question to Christoph concerning the nature of the devices used in his talk (the AR/VR components). VTK and openVR were used in much of the work regarding VR, Unity on AR side.

There was then a question to Remco concerning his slide about crystals. In the original diagram, there was a 'blank' region.  These regions may actually not have any actual (realizable) molecules in those blank regions.  It is not that there are always interesting molecules there; it might be, in fact, that no molecules exist there. It was acknowledged that physics-informed (or chemistry-informed) methods might allow one to distinguish between valid and invalid parts of the space.

The next question was concerning available toolkits for building VA systems.  It was acknowledged that there is not generally available toolkit available.

\clearpage
\SHONANabstract{Panel Discussion 5 (Thursday, 16 May 2024, morning session)}{Daisuke Sakurai, Kristi Potter, Tobias Schreck, and Christina Stoiber
}

This panel consisted of the following four speakers: Daisuke Sakurai, Kristi Potter, Tobias Schreck, and Christina Stoiber. Daisuke started the session off with a talk on multiple variables and interpretability. 
Kristi then presented an overview of work at NREL (US National Renewable Energy Lab) concerning uncertainty visualization.  
Tobias presented tools for visual analysis of production and test data.
Lastly, Christina presented a talk on visualization literacy and education.

The first question was to Christina concerning the use of 'comics'.  Do teachers use templates, or free-form?  They had a build-in set of predesigned templates to aid the teacher, but there is still flexibility.  People in the audience suggested that new generative models could be used to allow the user to prompt the generation of new items within the framework.

As a question concerning the use of comics for visualization onboarding: given that you have to explain the visualization to the user, explaining the data to the user, and explaining the narrative frame to the user, what are your experiences in finding these techniques to be most impactful? Current work is being done to see types of problems are best solved by the comic modality.  The current feedback from the users is that they like the comic paradigm.  There is still more work to be done as to the benefits for data, visualization, etc.

To Tobias, there was a question concerning outlier detection. The idea of outlier assumes that you have some idea of the distribution. The tool starts with creating a concept of what is 'normal', which in some sense gives a sense of what is outside 'normal'. It is a joint process of data analysis to get normal trends and then looping back to try to figure out what is abnormal. The audience member challenges that in some cases, the domain scientist knows quite clearly what is normal and abnormal (e.g., nuclear reaction). It was acknowledged that some applications are better suited by distributions and not fixed threshholds.

To Kristi, it was asked about dealing with uncertainty when the user does not really want to believe that there is uncertainty in the world.  Kristi acknowledges that she normally works with simulation scientists who acknowledge uncertainty. However, she acknowledges that in some domains, like policy, it is very important to understand both how you talk about and how you show uncertainty.

For Tobias, decision trees are very much appreciated by engineers.  Is this because of it being used in the auto engineers, or is there something more broad to be learned?  Decision trees are appreciated because they can be mapped to rules that can be written down (which helps with documentation).  Tobias did not look into the precision of this and how things translates to the rules that one would write down. An audience member said that there is a connection to predicate learning that could be explored.  Tobias pointed that that they often did several decision trees and then created summaries of the trees to help provide confidence in the results. Daisuke chimed in that, in his experience, people prefer decision trees as it helps give an idea of concrete rules that they can follow.

A question concerning Reeb-products: could this new technology be used for studying the roots of plants as shown in the morning lecture? There seems to be some possibility but it should be discussed offline.

\clearpage

\SHONANabstract{Panel Discussion 6 (Thursday, 16 May 2024, afternoon session)}{Renata Raidou, Wolfgang Aigner, Anja Helm, and Mennatallah El-Assady}

This panel consisted of the following four speakers: Renata Raidou, Wolfgang Aigner, Anja Helm, and Mennatallah El-Assady. Renata started the session off with a talk on harnessing visual computing in materials sciences through instights from biomedical visualization.
Wolfgang then presented visualization of time-oriented data. 
Anja presented visual comparisons of distributions of material data.
Lastly, Mennatallah presented co-adaptive analytics and guidance. 

Question concerning timeline revisited (paper by T. Munzner) addressed to Wofgang: can you speak to that simiplification and your thoughts.  Wolfgang acknowledged the work and said it should be considered a more specific case than the work that he presented. The question is whether Wolfgang's work explains Tamara's work.

For Menna, are there a set of optimization algorithms inside her framework. The current framework has very concrete (explicit) steps within the optimization process. In the future, the system should be adapted to learn the best strategies. In principle, the framework is flexible and the optimization components could be learned.  There was then a follow up discussion of the human-in-the-loop parts and the automated parts.  The current framework has a form of triage to try to up-front automate things first.

As a general question to all four: what is an example in each of your talks in which you can point to an application and tell us what we can learn for material science. Renata acknowledges that in biomedical sciences, there are lots of similarities to materials science in that there is multi-scalar data, time-dependent data, etc. Alpha-fold was mentioned by the audience as an example where something totally changed certain types of PhD items. Based upon that example: do any of your tools allow us to scan parameter spaces so much better in a way that will transform an applied domain. The panel argued that they are providing tools that enable the scientists, not replace them. Wolfgang highlighted the analogous use-cases in medicine as examples that show how these tools help applied areas.

One conclusion that came out is that we (materials science) might need to come up with a challenge problem with the data, metrics, etc., clearly stated so that the full weight of the computational community might be brought to bear.

Wolfgang asked Menna about agents and humans working together: is it multiple agents and multiple humans?  What do you do if there are conflicting opinions? Menna did clarify that they did not allow multi-human interaction in what she showed, but that there are trust models that can be used to help model and engage with multi-agent conflicting situations. There is a lot more that needs to be done in such scenarios.

For Anja, changing bin size can be challenging. Have you thought about automatically finding the parameters in your tool in an automatic way?  Anja acknowledged that such a question is, indeed, a natural follow-up problem that is to be addressed. 

To Renata: you mentioned sustainability in visualization.  However, many people are implementing (or re-implementing) various.  The scientific visualization community that you are engaging seems to have such specialization that you have to build from scratch each time.  Renata acknowledged that -- in an ideal case -- you could try to generate some of these collective tools. However, practically, we end up developing our own tools and then re-using the prototypes if they can move from PhD student to the next PhD student.  The reality is that different applications have different requirements, so lots of re-building often has to be done.

Menna asked of Wolfgang where there are other features that can be mapped to audio.  Wolfgang acknowledged that they are investigating the cases where things that are not best shown visually are specifically encoded through audio channels. There are other things than just pitch that can used to sensitize a human's audio channel. An audience member asked if you can change the direction of inference: is there music visualization?  Wolfgang acknowledged that there is an area within Visualization, but that he has not investigated it.

\clearpage

\section*{Summary of New Findings and Actionable Points}

As main actionable point the implementation of challenges and benchmarks has been identified. In the following sections these challenges and benchmarks are explained.

\subsection*{The Material Discovery Challenge:}
Starting point: in essence we started off with 3D Ashby diagram as shown by Guillermo. This visualization was drilled down to the actual data available from workflow. We figured out the following inputs and outputs.\\

Inputs:
\begin{itemize}
    \item $\alpha$: alloy components (50 - 60 elements, typically given alloy in wt.\%)
    \item $\beta$: process parameters (10 - 20 parameters)
    \item $\gamma$: model parameters (unknown number of parameters, model might be incomplete)
\end{itemize}

Data on material candidates is being generated using material simulation as well as the characterization of real material using imaging techniques.\\ 
\\
S = simulation\\
E = experiment\\
\begin{itemize}
    \item $S(\alpha, \beta, \gamma) \to S_o \to \text{QoI}_S$ (appr. 10 Quantities of interest determined in the simulation)\\
    \item $E(\alpha, \beta) \to E_o \to \text{QoI}_E$ (appr. 4 quantities of interest evaluated in the experiment)\\
\end{itemize}

\begin{figure}
    \centering
    \includegraphics[angle=180,origin=c,width=0.95\linewidth]{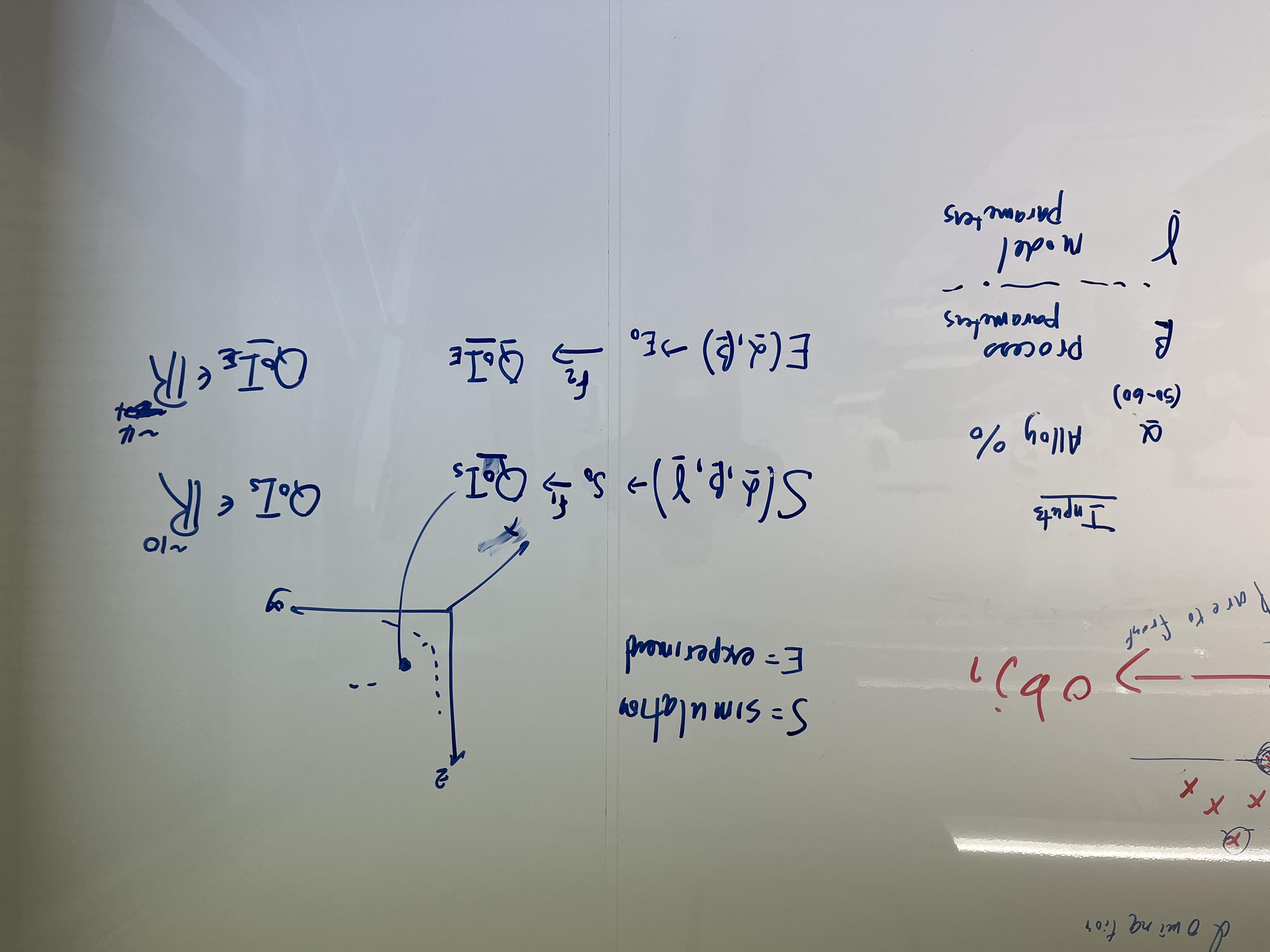}
    \caption{Material Discovery Challenge}
    \label{fig:MatDiscoChallenge1}
\end{figure}

\autoref{fig:MatDiscoChallenge1} shows the generated sketch for this challenge.
The application problem description is considered as the forward solver of the following general problem formulation: Multi objective optimization (see \autoref{fig:MultiObjOpt}): 
\begin{figure}
    \centering
    \includegraphics[angle=180,origin=c,width=0.95\linewidth]{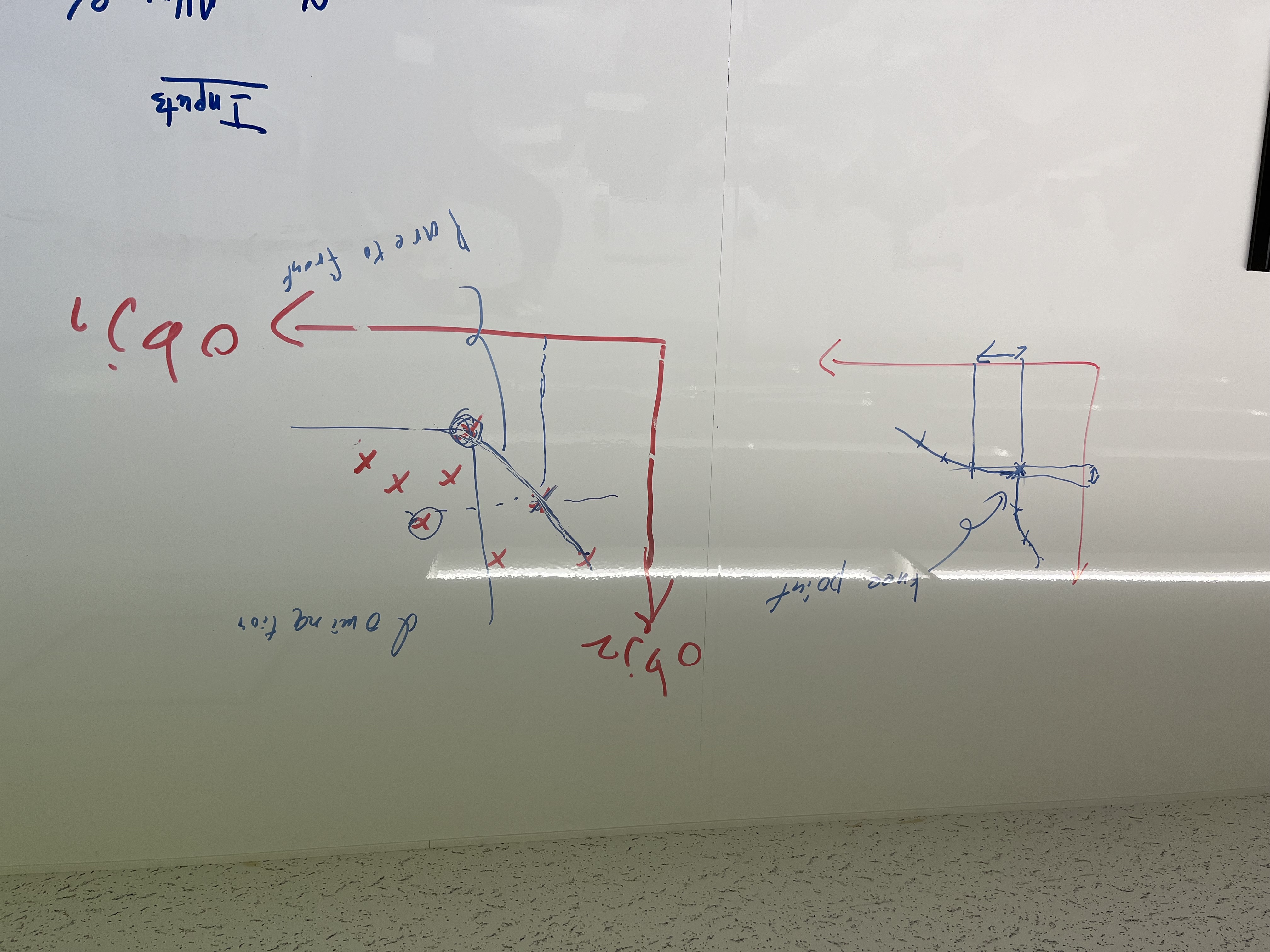}
    \caption{Multi objective optimization}
    \label{fig:MultiObjOpt}
\end{figure}
Multiple objectives, all cannot be considered at once. 
Simulation outputs dominate others $\to$ Domination
Pareto front exploration problem.
2D scatter plots are used for exploring but these do not give the high dimensional sense required for exploration.
Important in this consideration is the "knee point" analysis, which means, that gains on one axis come as small loss on the other axis (see figure).
In terms of the aforementioned application there, there is also a sensitivity problem to be considered. 
Small additions of a specific elements (in the range of ppm) can make the complete simulation data obsolete.

The described materials science problem seems to be suitable at least for the following areas in visualization research:
\begin{itemize}
    \item topological data analysis,
    \item sensitivity analysis,
    \item uncertainty analysis,
    \item optimization
\end{itemize}

\subsection*{The Material Characterization Challenge:}
The material analysis challenge: 
The discussed typical analysis workflow integrates X-ray imaging regarding a temporal exploration of fiber reinforced materials, 
i.e., heterogeneous materials integrating continuous unidirectional carbon fibers throughout the sample, matrix material as well as defects and inclusions
Fibers are not to be considered as uniform in terms of their structural parameters and thus strength, although strength is normalized regarding the diameter. 
If we test 100 fibers they will have a different strength which yields a strength distribution.
So, the micro-structure of the fibers in the dataset could have an impact on strength. A correlation of these aspects would be of interest.
Fiber breakages are key as they initiate the final fracture of the material. 
Neighborhood is potentially of interest (i.e., are fibers always at the same position or are they twisted, relocated etc.).
The phenomena we are looking into are very local.

Research questions are focusing on prediction from a domain perspective:
\begin{itemize}
    \item Are fiber breaks dominated by / related to micro-structure only?
    \item Can we predict the pattern of fiber breaks by just looking at the micro-structure?
\end{itemize}

Given the density of fibers and the strength distribution -> is there a relationship between those two, 
e.g., if we have more fibers in an area of interest, do we also have more breakages? Is there a probability of defects we can derive in a certain area?

Datasets being available on similar material types
\begin{itemize}
    \item 20 test series of unprocessed XCT data from 0N load to a special loading until final fracture (the last datasets include fiber breakages).
    \item 1-2 fully quantified test series.
\end{itemize}

The problem seems to be not suitable in this extent for a vis contest. Regarding ares in visualization research, these would be of interest:
\begin{itemize}
    \item Visual analysis of neighborhoods
\end{itemize}

\newpage

\section*{Identified Issues and Future directions} 

Possible outcomes from the workshop (identified in the Wednesday morning recapitulation session). These items have been extended and refined in the Thursday morning recapitulation. We structure them below in general and specific items:\\

\noindent
\textbf{General items:}
\begin{itemize}
    \item Dissemination statement for our seminar:\\
    Christoph Heinzl, Renata Georgia Raidou, Kristi Potter, Yuriko Takeshima, Mike Kirby, Guillermo Requena. ``Advancing Visual Computing in Materials Sciences" (Shonan Seminar 189). Shonan Seminar Report 189, pp. 1-63, Shonan Village Center, NII National Institute of Informatics, (05/2024)\\   
    https://shonan.nii.ac.jp/seminars/189/
    \item Another Dagstuhl and/or Shonan. (Interested: Gerik Scheuermann is willing to help, Christoph Heinzl, Marco Angelini in participating)
    \item Writing joint papers (white papers), position papers, books.
    \item Benchmarking papers (common data set and get different perspectives / approaches).
    \item Consider a book.
    \item Consider donating your brain. 
\end{itemize}

\noindent
\textbf{Specific Items:}
\begin{itemize}
    \item Vis Contest (Scientific Visualization Contest): Benchmark from some application domain. Get together a group of visualization and application scientists to set out data and tasks, and see how various visualization groups engage on the tasks. (Interested: Christoph Garth, Christoph Heinzl, Guillermo Requena, Mike Kirby, Gerik Scheuermann, Renata Raidou) 

    \item Application Spotlight at the IEEE VIS conference in Tampa, FL, USA, 13-18 October, 2024.  Application Spotlight organizers will be asked to provide a title along with a summary of the spotlight (no more than 500 words) describing the topic area, basic research questions, and why/how VIS technology can create meaningful benefits in that area. Deadline for the proposal is June 6th. The CFP:\\  \href{https://ieeevis.org/year/2024/info/call-participation/application-spotlights}{https://ieeevis.org/year/2023/info/call-participation/application-spotlights}. (contact point: Kristi Potter, Dani Ushizima. Interested: Christina Stoiber, Marco Angelini, Christoph Heinzl).

    \item Workshop at VIS. Workshops provide an informal setting in IEEE VIS for participants to discuss advanced topics in visualization, involve experts in the field, disseminate work in progress, and promote new ideas. Organizers will be asked to submit a 4-page paper including logistics and organization details, the planned activities including an outline of the schedule, the intended impact and a brief justification for why this workshop is necessary. Deadline for 2024 has passed, but will be sometime in February for Vis '25 (in Vienna). 2024's CFP:\\ \href{https://ieeevis.org/year/2024/info/call-participation/workshops}{https://ieeevis.org/year/2024/info/call-participation/workshops}. (Interested: Renata Raidou).

    \item Probably writing a joint challenge paper in terms of combining material properties simulations with pruning techniques to get faster to a result. Imaging using machine learning to identify a region within the search space and doing the physically accurate simulations only therein. (Interested: Marco Angelini, Kristi Potter, Renata Raidou)

    \item Granular Gordon Conference: Good starting point where people publish unpublished data. \href{https://www.grc.org/granular-matter-conference/2024/}{https://www.grc.org/granular-matter-conference/2024/}. (Contact: Matthias Sperl)

    \item MSE Conference and EuroMAT. Create a symposia for that conference. Guillermo could be the point person. (Contact: Guillermo Requena)

    \item Visualization Viewpoint article that gives challenges at the interface of VIS and Materials Science. (Interested: Velitchko Filipov, Christoph Heinzl, Marco Angelini, Christina Stoiber, Gerik Scheuermann, Kristi Potter, Renata Raidou,
    Anja Heim)

    \item Special Issue of Journal of Imaging or CG\&A or Computers and Graphics. (Interested: Velitchko Filipov, Christoph Heinzl, Renata Raidou)

    \item Joint proposals (open call OSCAR). Marie Curie. Possible joint calls such as US NSF with DFG (Germany), WEAVE (bilateral cooperative projects), EPSRC (Britain). (Interested: Christoph Heinzl, Gerik Scheuermann, Hamish Carr, Renata Raidou)

    \item Secondments, i.e., sending VIZ students to the materials science institutions (DLR would be fine with that, KU Leuven most likely also) (Interested: Christoph Heinzl, Gerik Scheuermann, Renata Raidou)

    \item Invited Talks, from Visual Computing to Materials Science and vice versa (Interested: Christoph Heinzl)

    \item Joint zoom call for further communication, about in half a year. In addition, a joint discord channel or similar could be used (e.g., Slack, which allows the centralization of materials and communications). (Interested: Velitchko Filipov, Christoph Heinzl (organizer), Marco Angelini, Gerik Scheuermann, Kristi Potter, Renata Raidou, Wolfgang Aigner)

    \item Challenges / Benchmark studies: Discussed areas were improvement / support in material discovery; 
    \begin{itemize}
        \item The Material Discovery Challenge: Challenge improving Ashby Diagram
        \item The Material Analysis Challenge: Benchmark on materials analysis pipeline
    \end{itemize}
    (Interested: Velitchko Filipov, Wolfgang Aigner, Christoph Heinzl, Marco Angelini, Christina Stoiber, Renata Raidou,
    Anja Heim)
    
\end{itemize}

\clearpage
\bibliographystyle{abbrv}
\bibliography{references}

\end{document}